\documentclass[fleqn,usenatbib]{mnras}

\usepackage{newtxtext,newtxmath}

\usepackage[T1]{fontenc}

\DeclareRobustCommand{\VAN}[3]{#2}
\let\VANthebibliography\thebibliography
\def\thebibliography{\DeclareRobustCommand{\VAN}[3]{##3}\VANthebibliography}

\usepackage{graphicx}	
\usepackage{amsmath}	

\usepackage{microtype} 

\usepackage{etoolbox, siunitx}
\robustify\bfseries

\usepackage{booktabs} 

\usepackage[normalem]{ulem} 

\DeclareSIUnit\au{AU}
\DeclareSIUnit\Rsun{R_\odot}
\DeclareSIUnit\Rjup{R_\text{Jup}}
\DeclareSIUnit\Msun{M_\odot}
\DeclareSIUnit\Mjup{M_\text{Jup}}
\DeclareSIUnit\gyr{Gyr}
\DeclareSIUnit\ppt{ppt}
\DeclareSIUnit\ppm{ppm}

\title[True Unicorns and False Positives]{True Unicorns and False Positives: Simulated Probabilities of Dark Massive Companions to Bright Stars}

\author[Miller et al.]{%
        Andrew M. Miller$^{1,2}$$^{\href{https://orcid.org/0009-0009-0496-277X}{\includegraphics[scale=0.5]{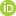}}}$,
        Alexander P. Stephan$^{3,2,4}$$^{\href{https://orcid.org/0000-0001-8220-0548}{\includegraphics[scale=0.5]{orcid.jpg}}}$ \&
        David V. Martin$^{5,2,6}$$^{\href{https://orcid.org/0000-0002-7595-6360}{\includegraphics[scale=0.5]{orcid.jpg}}}$
\vspace{0.3cm}
\\
\vspace{0.3cm}
andrew.miller21@rockets.utoledo.edu\\
$^{1}$Department of Physics and Astronomy, The University of Toledo, Toledo, OH 43606, USA\\
$^{2}$Department of Astronomy, The Ohio State University, 4006 McPherson Laboratory, Columbus, OH 43210, USA\\
$^{3}$Department of Physics \& Astronomy, Vanderbilt University, Nashville, TN 37235, USA \\
$^{4}$Center for Cosmology and AstroParticle Physics, The Ohio State
University, Columbus, OH 43210, USA \\
$^{5}$Department of Physics \& Astronomy, Tufts University, Medford, MA 02155, USA \\
$^{6}$NASA Sagan Fellow\\
}

\vspace{-2cm}
\date{First submitted to MNRAS on March 3, 2023}

\pubyear{2023}

\begin{document}

\label{firstpage}
\pagerange{\pageref{firstpage}--\pageref{lastpage}}
\maketitle

\begin{abstract}

Many compact objects (black holes and neutron stars) exist in binaries. These binaries are normally discovered through their interactions, either from accretion as an X-ray binary or collisions as a gravitational wave source. However, the majority of compact objects in binaries should be non-interacting. Recently proposed discoveries have used radial velocities of a bright star (main sequence or evolved) that are indicative of a massive but dark companion, which is inferred to be a compact object. Unfortunately, this burgeoning new field has been hindered by false positives, including the ``Unicorn'' (V723 Mon) which was initially believed to be a red giant/black hole binary before being refuted. In this work, we investigate the evolution of stellar binary populations over time, using the binary evolution code COSMIC to simulate binary populations and determine the probability of a candidate object being either a ``true Unicorn'' (actual compact objects in binaries) or a false positive. We find that main sequence stars have a higher true Unicorn probability than red giants or naked helium stars (an exposed core of an evolved star), particularly if the companion is more massive and is $\ge$3 times less luminous than the MS star. We also find that a top-heavy initial mass function raises the true Unicorn probability further, that super-solar metallicity reduces the probability, and that most true Unicorns are found at periods $\le$100 days. Finally, we find that a significant fraction of true Unicorns do not evolve into x-ray binaries during the age of the universe.

\end{abstract}




\section{INTRODUCTION}\label{sec:introduction}

Historically, neutron stars and stellar mass black holes have been discovered in interacting binaries. These interacting objects include X-ray binaries, where material from a donor star is accreted onto a compact object, emitting X-rays, \citep{Giacconi1964,Overbeck1968,Shakura1973}, and more recently, the merger of two compact objects and the production of gravitational waves \citep{Abbott2016,ligo2021}. Compact objects can also exist in non-interacting binaries \citep{Guseinov1966,Trimble1969}. There are two primary motivations to search for non-interacting compact objects: 1) it is expected that the majority of compact objects live in non-interacting binaries \citep{Mashian2017}, and 2) solely observing interacting binaries provides a biased sample of compact objects, distorting the observed compact object population and hindering our understanding of the processes which produced them. 

Compact objects in non-interacting binaries can be found in several ways, including using pulsars which are eclipsed \citep{Yan2021} or have timing variations \citep{Lorimier2021}. Astrometry is another method, and the final \textit{Gaia} data release should contain hundreds or thousands of compact objects in non-interacting binaries \citep{Mashian2017,Yamaguchi2018, Wiktorowicz2020, Chawla2022}. \textit{Gaia} may have the potential to characterize every black hole with a luminous companion located within 1 kpc \citep{Andrews2019}. However, if \textit{Gaia} data on the luminous star is incomplete, then only the total mass of the system is known, and additional methods are needed to break the binary's mass degeneracy and determine the companion's true mass.

A technique which has found popularity recently is using radial velocities (RVs). In this technique, RVs indicate a binary pair with two stellar mass objects, however both the spectroscopy and spectral energy distribution (SED) are consistent with only a single ``bright'' star. Based on the masses of the objects, it is argued that if the companion were a regular star, then it would be visible. Therefore, the companion must be massive but ``dark'', and hence a compact object (neutron star or black hole).

\citet{Thompson2019} studied the red giant 2MASS J05215658+435922. ASAS-SN photometry showed a roughly sinusoidal 83.2-day periodicity, which was attributed to star spots on a rotating star. Initial APOGEE radial velocities suggested a massive companion. Follow-up with the TRES spectrograph confirmed a dark companion with a minimum mass of $2.9$~M$_\odot$, comparable to the mass of the bright giant ($3.2$~M$_\odot$). The fact that an only slightly less massive companion was not visible was evidence for a compact object. There was also an absence of X-ray emission, and hence this was deduced to be a black hole in a non-interacting binary. The observations do not directly distinguish between a black hole and a neutron star for the dark companion, however a black hole was favoured since the conventional upper mass limit for a neutron star is $2.9-3.0$~M$_\odot$ \citep{Finn1994, Kalogera1996}. At $\ge2.9$~M$_\odot$, this would be the lowest mass black hole discovered, placing it in the ``mass-gap'', an interesting region of parameter space in between neutron stars and black holes. Currently, this discovery has not been overturned, and is still believed to be a black hole.

More recently, \textit{Gaia} astrometric measurements were combined with radial velocity and spectroscopic follow-up from multiple instruments to discover Gaia BH1, a main sequence G star orbiting a $\sim$10 M$_\odot$ black hole \citep{El-Badry2023}, as well as Gaia BH2, a $\sim$1 M$_\odot$ red giant orbiting a $\sim$9 M$_\odot$ black hole \citep{El-Badry2023RG}. At 480 pc, Gaia BH1 represents the nearest black hole discovered to date. These were followed by Gaia NS1, a main sequence G star orbiting a $\sim$1.9 M$_\odot$ neutron star \citep{ElBadry2024NS1}, and Gaia BH3, a giant-branch G star on an 11.6 year orbit with a $\sim$32.7 M$_\odot$ black hole \citep{GaiaCollab2024}, which at 590 pc is the second-nearest black hole discovered. These methods were also used to identify a population of 21 main sequence stars orbiting neutron star candidates with masses between $1.261$ and $1.898$ M$_\odot$ \citep{ElBadry2024NScand}. For some of these candidates, a white dwarf has not been ruled out due to the mass degeneracy between high-mass white dwarfs and low-mass neutron stars.

While these methods have been successful, they have also resulted in false positives. The \citet{Thompson2019} discovery was followed by similar discoveries in V723 Mon (aka ``the Unicorn'') by \citet{Jayasinghe2021Unicorn} and 2M04123153+6738486 (aka ``the Giraffe'') by \citet{Jayasinghe2022Giraffe}, using comparable methods and datasets. As in \citet{Thompson2019}, both targets have no visible X-ray emission, although \citet{Jayasinghe2021Unicorn} did invoke an accretion disc around V723 Mon's black hole to explain the observed spectral energy distribution (SED). These were also proposed as mass-gap black holes in non-interacting binaries. Shortly after, \citet{ElBadry2022BinaryZoo} refuted the presence of a black hole in both binaries. A re-analysis of the SED and spectroscopy through spectral disentangling showed that the companions, whilst faint, are not completely dark, and hence are not compact objects. For V723 Mon, the companion is particularly nefarious, since it is a rapidly-rotating subgiant, where the spectral lines are so heavily broadened that they become effectively invisible.

Our Galaxy should contain as many as 10,000 stellar/compact object binaries \citep{Chawla2022}. However, the number of objects which could impersonate these binaries in terms of mass and luminosity is unknown. Since this burgeoning field of non-interacting binary discovery has faced early challenges, we seek to investigate the probability that a compact object candidate, once it has been identified as part of a sky survey, is either a genuine compact object or an imposter star.  We define non-interacting binaries containing an actual compact object as ``true Unicorns''. While in \citet{Jayasinghe2021Unicorn}, the ``Unicorn'' name was applied specifically to V723 Mon, in this study we apply it broadly to all non-interacting binaries containing one star and one compact object, a class of objects which V723 Mon was believed to be before it was refuted. We define a ``false positive'' as a stellar binary whose SED appears to indicate only a single star, however spectral disentangling would reveal the presence of a second, less luminous star. If this companion is extremely faint, its contribution to the SED can be virtually undetectable \citep{El-Badry2018}, requiring significant effort to classify the binary. We seek to analyze the types of stars and stellar populations for which true Unicorns are likely to outnumber false positives, and to provide criteria that will give observers a higher probability of identifying binary systems containing a non-interacting compact object.

For this purpose, we simulate stellar binary populations and track their evolution for 13 Gyrs. We first explore a generic, high-mass Galactic population. We then explore a high-mass population similar to what would be found in the Arches Cluster, a dense stellar cluster toward the central region of our galaxy with a young age of $\sim3.7\pm0.2$~Myrs \citep{Hosek2019,Gallego2021}. This cluster is believed to have a top-heavy initial mass function \citep{Hosek2019} as well as super-solar metallicity \citep{Martins2008, Sabhahit2022}, which we predict will produce a higher probability of compact objects compared to false positives. In Section \ref{sec:methodology} we outline the physical parameters of our binary populations and the code used to evolve them, and present our general results in Section \ref{sec:discussion} before concluding in Section \ref{sec:conclusion}.

\section{Methodology}\label{sec:methodology}

\subsection{Binary Population Synthesis}\label{subsec:mathematical_methods}

We construct a population of  binaries based on  literature distributions for four key parameters: initial mass function (IMF), mass ratio $q=m_{2}/m_{1}$, orbital period $P_{\rm bin}$, and orbital eccentricity $e_{\rm bin}$. 

The IMF is generally described by the power law 
$$
\xi(m) \propto {m^{-\alpha}},
$$
where we use two values for $\alpha$ in order to compare two separate varieties of stellar populations in our galaxy. We use $\alpha_K = 2.7$ to model the IMF of the solar neighborhood and overall galactic star population, based on \citet{Kroupa1993} \citep[see also][]{Salpeter1955}. We use $\alpha_H = 1.8$ based on \citet{Hosek2019} \citep[see also][]{Kim2006}, which describes the IMF of the Arches cluster. This second IMF is top-heavy compared to the general Kroupa IMF, producing a higher number of massive stars. Through this, we examine whether young, dense clusters near the Galactic center might be home to more binaries containing non-interacting compact objects. We also probe the general effect caused by a change in IMF. The $\alpha$ values we reference here are valid for initial primary stellar masses above $1$~M$_\odot$.

We model the values of $q$, $\log P_{\rm bin}$, and $e_{\rm bin}$ as Gaussian distributions based on the results of \citet{Duquennoy1991}. The distribution parameters, such as the average expected value $\mu$ and the standard deviation $\sigma$, are listed in Table \ref{tab:table1}. The eccentricity distribution is valid for binary periods less than about 1000 days, which covers most of the period range focused on in our study (see Section \ref{subsec:modelling}).

\begin{table}
  \begin{center}
  \caption{Distribution parameters for the variables used in constructing our binary populations, namely primary star mass $m_1$, mass ratio $q$, orbital period $P_{\rm bin}$, and orbital eccentricity $e_{\rm bin}$, based on \citet{Duquennoy1991}. In particular, we give the $\alpha$ values for the IMFs based on \citet{Kroupa1993} and \citet{Hosek2019}, as well as expected value $\mu$, and standard deviation $\sigma$ for the Gaussian distributions. We also give the minimum and maximum variable values.}
  \label{tab:table1}
    \begin{tabular}{c|c|c|c|c|c|c}
      Variable & Function & $\alpha_K$ & $\alpha_H$ & Min & Max & Units\\
      \hline
      $m_1$ & Power Law & 2.7 & 1.8 & 2 & 150 & M$_\odot$\\
      \hline
      \hline
      Variable & Function & $\mu$ & $\sigma$ & Min & Max & Units\\
      \hline
      $\log P_{\rm bin}$ & Gaussian & 4.8 & 2.3 & 0 & 8.56 & $\log {\rm Days}$\\
      $q$ & Gaussian & 0.23 & 0.42 & 0.001 & 0.999\\
      $e_{\rm bin}$ & Gaussian & 0.31 & 0.35 & 0.0 & 1.0\\
      
    \end{tabular}
  \end{center}
\end{table}

Using the distributions described in Table \ref{tab:table1}, we construct three separate populations of $5\times10^{5}$ binary pairs each, one employing the IMF of \citet{Kroupa1993}, the second based on the IMF of \citet{Hosek2019}, and the third based on the Hosek IMF with super-solar metallicity as per \citet{Martins2008} and \citet{Sabhahit2022}. We choose $2$~M$_\odot$ as the minimum mass at formation for the primary star of each binary, as we are  interested in relatively massive binary pairs that could either produce or mimic a hidden compact object. We treat each population as a cluster with all stars roughly equidistant from Earth, allowing us to focus on objects' luminosity instead of flux.

\subsection{Measuring Radial Velocities}\label{subsec:measuringradialvelocities}

To ensure that the binaries we examine could truly be detected through radial velocity measurement, we calculate the radial velocity semi-amplitude of the more luminous star in each binary pair.

To calculate the RV semi-amplitude, we use:
$$
K_{\rm Bright} = \left( \frac{2 \pi G}{P_{\rm orb}} \right)^{1/3}\frac{m_{\rm Bright} \sin{i}}{{\left(m_{\rm Bright}+m_{\rm Dark}\right)}^{2/3}}\frac{1}{\sqrt{1-e^{2}}}
$$
where $m_{\rm Bright}$ is the mass of the bright star in each binary, and $m_{\rm Dark}$ is the mass of the less luminous companion.

For $\sin{i}$, we assign the value $1/\sqrt{2}$ to all binary pairs. This represents an average inclination within a stellar population, and is consistent with the expectation that lower inclinations (<30$^{\circ}$) are geometrically disfavored in binaries \citep{El-Badry2022BHImposter}.

\subsection{Identifying X-Ray Binaries}\label{subsec:xraybinaries}

As the focus of our study is to identify the number of binary systems where object classification could be open to misinterpretation, we seek to remove x-ray binaries from our population sample. X-ray binaries provide direct evidence of compact objects through the presence of an accretion disk and detectable x-ray radiation. This differentiates them from most stellar binaries that could act as false positives. To remove them, we must first define which objects within our populations could be detected through x-ray emission at different times.

High-mass x-ray binaries (HMXBs) are predominantly formed when O and A type stars fill at least 80 to 90 percent of their Roche lobe, creating a wind-driven accretion stream onto the surface of a compact object companion \citep{Hirai2021}. The stars typically have mass $\ge10$~M$_\odot$ \citep{Corral-Santana2015}. During overflow they lose mass at a rate approaching $10^{-6}$~M$_\odot$ yr$^{-1}$ \citep{King1995}. These objects are rare in a galaxy the size of the Milky Way, where we should see only 1-2 observable black hole HMXBs at a time \citep{Romero-Shaw2023}.

Low-mass x-ray binaries (LMXBs) are formed when stars with mass $\le1$~M$_\odot$ transfer material via Roche lobe overflow through their inner (L1) Lagrangian point onto the surface of a compact object companion \citep{Charles&Coe2006}. Objects detected as persistent LMXBs lose mass at a rate near $10^{-8}$~M$_\odot$ yr$^{-1}$, while transient LMXBs lose mass at a rate near $10^{-9}$~M$_\odot$ yr$^{-1}$\citep{Tanaka&Shibazaki1996}.

While intermediate-mass x-ray binaries (IMXBs) have been discovered \citep[e.g.,][]{Tananbaum1972} and should be forming at a rate $\ge$5 times that of LMXBs \citep{Pfahl2003}, relatively few have been identified within the Galaxy to date \citep{Hunt2021}. IMXBs tend to quickly evolve into LMXBs through mass loss, making them difficult to characterize \citep{Podsiadlowski2002, Pfahl2003}. Their mass loss rates are also not well understood, given that stars of 1-10 M$_\odot$ are typically not large enough to generate the high wind-driven loss needed to produce an observable x-ray source \citep{Tauris2006}.

To identify x-ray binaries, we adopt the methodologies of \citet{Podsiadlowski2003}, \citet{Liotine2023}, and \citet{Misra2023} to calculate the x-ray luminosity of each BH and NS undergoing accretion within our populations. The x-ray luminosity of the compact object is defined by
$$
L_{\rm x} = \eta \dot{M}_{\rm acc} c^{2}
$$
where $\dot{M}_{\rm acc}$ is the mass-accretion rate and $c$ is the speed of light. Since not all accreting material will contribute to the compact object's growth, $\dot{M}_{\rm acc}$ is related to the actual change in the compact object's mass ($\dot{M}_{\rm CO}$) by
$$
\dot{M}_{\rm acc} = \frac{\dot{M}_{\rm CO}}{1-\eta}
$$
The radiative efficiency, $\eta$, is the fraction of accreted rest mass which is converted into energy and radiated away, and is defined as
$$
\eta = \frac{G M_{\rm CO}}{R_{\rm acc} c^{2}}
$$
where $G$ is the gravitational constant, and $M_{\rm CO}$ is the mass of the compact object. For neutron stars, $R_{\rm acc}$ is simply the object's radius, while for black holes, $R_{\rm acc}$ is the spin-dependent innermost stable circular orbit around the object \citep{Podsiadlowski2003}, defined as
$$
R_{\rm ISCO} = 6 \frac{G M_{\rm BH}}{c^{2}}
$$
We define x-ray binaries (XRBs) as any binary containing a non-degenerate star and an accreting BH or NS with a calculated x-ray luminosity > 10$^{35}$ erg s$^{-1}$ \citep{Misra2023}. At this luminosity, we should be detecting both transient and persistent sources. One of the least luminous transient XRBs detected, XTE J1118+480, peaks at $\sim$3.6$\times$10$^{35}$ erg s$^{-1}$ \citep{Dunn2010, Corral-Santana2015}. Our sample may not include low luminosity Be XRBs \citep{Pfahl2002}, which show persistent x-ray luminosity in the range of 10$^{34}$-10$^{35}$ erg s$^{-1}$ with bursts $\sim$10 times higher \citep{Sguera2023}. 

We further classify our XRB population based on the bright star's mass: HMXBs, with masses $\ge10$ M$\odot$; LMXBs, with masses $\le1$ M$\odot$; and IMXBs, with masses from 1-10 M$\odot.$

\subsection{Modelling}\label{subsec:modelling}

The binary populations we create with the parameters outlined in Section \ref{subsec:mathematical_methods} are evolved using {\tt COSMIC} \citep{Breivik2020}. {\tt COSMIC} is a binary population synthesis and evolution code based on the older {\tt BSE} code \citep{Hurley2002}, however it greatly expands upon the functionalities of {\tt BSE} by including more detailed stellar and binary interaction processes and allowing for the evolution of more massive stars. {\tt COSMIC} has been a popular tool of late, including the recent \citet{Weller2023} predictions for Milky Way Mapper in SDSS-V, which searches for progenitors of compact object collisions that will produce gravitational waves, as well as \citet{Liotine2023}, which probes the lack of observed HMXBs that are predicted to evolve into binary black hole mergers. {\tt COSMIC} has also been used to probe \textit{Gaia's} ability to detect stellar/compact object binaries in the Galaxy \citep{Andrews2019, Breivik2019CO, Chawla2022}.

For our work we use {\tt COSMIC} version $3.4.0$ to evolve each binary for $13$~Gyrs. We use solar metallicity ($0.0134$) for each star \citep{Asplund2009} within the Kroupa and Hosek IMF populations. We also perform additional modelling for the Hosek IMF at higher metallicity (see Section \ref{subsec:ArchesHighMetallicity}). 

The general input physics parameters for {\tt COSMIC} are outlined in \citet{Breivik2020}. For this study we have used {\tt COSMIC}'s default values, representing reasonable parameters based on current understanding of stellar evolution. These values are listed in Appendix \ref{sec:appendixA}. Of note is the `remnantflag' parameter which determines the mass prescriptions for neutron stars and black holes, providing an option between \textit{rapid} and \textit{delayed} supernova explosion models. The mass distribution of black holes has been found to depend strongly on the choice of this parameter \citep{Chawla2022}. The \textit{rapid} model assumes a mass gap between neutron stars and black holes. \citet{Breivik2019CO} notes a decrease in the overall number of black holes when building populations under this model, and notes inconsistencies with observation. Therefore, we have employed the \textit{delayed} model, which assumes no mass gap and is the default in {\tt COSMIC}. 

\begin{table}
  \begin{center}
    \caption{List of evolutionary stellar states produced by {\tt COSMIC}, and the acronyms used in this study to refer to each group of states. {\tt COSMIC} also produces white dwarfs, which appear within our populations but are not listed here since they are outside the focus of this study.}
    \label{tab:table2}
    \begin{tabular}{c|c}
        Acronym & {\tt COSMIC} Description\\
        \hline
        MS & Main Sequence, <0.7 M$_\odot$\\
        & Main Sequence, >0.7 M$_\odot$\\
        \hline
        RGB & Hertzsprung Gap\\
        & First Giant Branch\\
        & Core Helium Burning\\
        & Early Asymptotic Giant Branch (AGB)\\
        & Thermally Pulsing AGB\\
        \hline
        NHS & Naked Helium Star MS\\
        & Naked Helium Star Hertzsprung Gap\\
        & Naked Helium Star Giant Branch\\
        \hline
        NS & Neutron Star\\
        \hline
        BH & Black Hole\\
    \end{tabular}
  \end{center}
\end{table} 

The output of our {\tt COSMIC} simulations consists of the evolutionary states of the binary pairs at a given time, as well as additional characteristics such as mass, orbital period, and effective temperature. Table \ref{tab:table2} shows the {\tt COSMIC} evolutionary states which we focus on in this study, as well as the acronyms used to refer to these states.

{\tt COSMIC} provides the bolometric luminosity of each star as an output at each time interval. When we refer to an object as ``bright'' or ``dark,'' we are referring to its total bolometric luminosity and not its appearance in visible wavelengths. In categorizing stars by their bolometric (as opposed to visible) luminosities, we are consistent with the methods used to further analyze candidate non-interacting compact objects once they are detected through visible-light surveys. The combined total luminosity of both stars in a binary determines the shape of the spectral energy distribution (SED), and in the case of faint companions, further analysis of the spectrum and SED are needed to confirm the presence of either two stars, or of one star and a compact object.

{\tt COSMIC} tracks binary evolution in two ways: 1) by sampling the binary characteristics each time a key evolutionary change occurs, including the exact time of the change; and 2) by sampling the binary parameters at a series of user-defined time steps, for which we used 37 logarithmic steps. We then combine these two different outputs and sort our binaries' evolution into a series of logarithmic time interval bins. Logarithmic bins were chosen as stellar populations on the whole tend to exhibit more change at early ages due to the rapid evolution of massive stars. At later ages, the population as a whole changes more slowly as lower-mass stars continue their gradual evolution. Combining the two {\tt COSMIC} outputs ensures that no significant evolutionary change is missed in our data.

Within each time interval bin, we track the number of \textit{true Unicorns}, which is the name we assign to non-interacting binaries containing one compact object, consistent with the class of binary postulated by \citet{Jayasinghe2021Unicorn}.  We also track potential \textit{false positives}, which are stellar binaries that mimic true Unicorns by exhibiting similar mass and luminosity characteristics, such as the objects identified in \citet{ElBadry2022BinaryZoo}, \citet{El-Badry2022BHImposter}, and \citet{El-Badry2022AnotherImposter}. Since false positives contain one star less luminous than the other, spectral disentangling should be needed to determine the presence of the ``dark'' star. These binaries would require additional effort and spectral analysis to categorize.

We seek to ensure that our count of true Unicorns and false positives reflects what an observer would find when viewing our stellar population at an instantaneous moment within a given time bin. Therefore, the count of objects within each time bin is weighted by the amount of time that object spends as a true Unicorn or false positive. For example, if the time bin's length is 1 Gyr, and the binary forms a false positive for only 10 Myrs within that bin, we count the binary as 0.01 false positives for that time bin. If the time bin's length is less than the amount of time the binary spends as a false positive (3 Myrs, for example), we count the object as 1 false positive for that time bin.

We employ a number of criteria to define which objects fall within the true Unicorn and false positive object classes during each time interval. Within each interval, we remove any binary where the less luminous object has mass <$1.4$~M$_\odot$, since we are primarily concerned with binaries that either contain a BH or NS (true Unicorns), or which contain a less luminous star that mimics the mass of a BH or NS (false positives). This means that some neutron stars are removed from our population, since there is a range of masses for which neutron stars overlap with white dwarfs. White dwarfs are the remnants of low-mass stars, with observed white dwarf masses ranging from $0.136-0.162$~M$_\odot$ \citep{Kawka2009, Kaplan2014} to $1.327-1.365$~M$_\odot$ \citep{Caiazzo2021}. Neutron stars form following supernovae when the compact object has a minimum post-supernova mass $\ge0.95-1.29$~M$_\odot$ \citep{Strobel1999, Lattimer2004}. While the minimum mass for a NS built of cold catalyzed matter can be much lower at $\sim$0.09 M$_\odot$ \citep{Haensel2002}, it is only possible to reach this size through the interaction of short-period NS-NS binaries \citep{Yudin2020}. 

By excluding any binaries where the less luminous object has mass <$1.4$~M$_\odot$, we avoid objects for which the classification between neutron star and white dwarf would be uncertain, as it is with some of the NS candidates reported in \citet{ElBadry2024NScand}. Within our populations, $\sim$65-70\% of low-luminosity NS fall below the mass cut and are removed. However, some low-mass neutron stars are still counted. $\sim$14-21\% of NS with mass <1.4 M$_{\odot}$ are hot, recently-formed objects (<100 Myrs in age) with higher bolometric luminosity than their stellar companions. In these binaries, the NS is the ``bright'' object in the SED while the stellar companion is the ``dark'' object, allowing the binary to be counted as a true Unicorn in spite of the low NS mass. Within our simulations, we find no white dwarfs with masses higher than $1.39$~M$_\odot$. Throughout this study, we use ``compact objects'' to refer to BH and NS only, since white dwarfs are not a focus of this work.

Within each time interval, we also remove any binary pair where the bright star has a radial velocity semi-amplitude \textit{K} < 1 km/s. While many instruments are capable of detecting smaller values, 1 km/s represents a conservative lower limit of \textit{Gaia} RV detection \citep{Jordan2008}, with typical \textit{Gaia} RV semi-amplitudes expected to range from several to hundreds of km/s \citep{Chawla2022}.  By excluding objects with \textit{K} < 1 km/s, we ensure that binaries used in our model would be detectable through large-scale surveys.

We also remove binaries with orbital periods >1826 days (5 years) within each time interval. Our {\tt COSMIC} output generates stars with orbital periods far too long to be observed in the timescale of human civilization which nonetheless have an RV semiamplitude of $K\ge1$ km/s. We seek to ensure that binaries in our sample have a short enough period that their RV semiamplitude can be calculated using a reasonable observing timeframe of a few years. Many of the star/compact object binaries discussed in Section \ref{sec:introduction} have periods below this limit, lending credence to our selection. 2MASS J05215658+435922 has an orbital period of 83.2 days \citep{Thompson2019}, Gaia BH1 has a period of 185.6 days \citep{El-Badry2023}, Gaia NS1 has a period of 731 days \citep{ElBadry2024NS1}, and Gaia BH2 has a period of 1277 days \citep{El-Badry2023RG}. The 21 neutron star candidates proposed in \citet{ElBadry2024NScand} have periods ranging from 189-1046 days. The high-mass Gaia BH3 is an outlier with an 11.6-year orbit \citep{GaiaCollab2024}. Most objects of interest are expected to have periods on order of a few hundred days, as there is an observed lower incidence of binary BH companions in \textit{Gaia} data with periods of 400-1000 days \citep{El-Badry2023}.

Within each time interval, we also exclude any binary pairs which have formed an active x-ray binary (LMXB, IMXB, or HMXB), since we are only interested in tracking the number of compact objects whose presence is not clearly shown through x-ray radiation at a given time. To identify x-ray binaries within {\tt COSMIC}, we measure the change in mass of each BH and NS between each time interval ($\dot{M}_{\rm CO}$), and use this to calculate the mass-accretion rate and x-ray luminosity using the equations in Section \ref{subsec:xraybinaries}.

Finally, we exclude all binaries that have either merged or been disrupted within each time interval, as noted by {\tt COSMIC}.

\subsection{True and False Positive Definitions}\label{subsec:truefalsepositive}

We define true Unicorns as a NS or BH orbited by a MS, RGB, or NHS (in {\tt COSMIC}, a naked helium star is defined as an evolved star severely stripped by mass loss during core helium burning, resulting in exposure of the nuclear processed material in the core \citep{Hurley2000}). False positives are defined as a pair of stars (MS/MS, RGB/MS, RGB/RGB, NHS/MS, or NHS/RGB) which mimic true Unicorns in mass and luminosity, such as those objects characterized in \citet{ElBadry2022BinaryZoo}, \citet{El-Badry2022BHImposter}, and \citet{El-Badry2022AnotherImposter}.

We examine the numbers of true Unicorns and false positives within our population under two different definitions:

\begin{itemize}
  \item {{\bf Criteria 1:} The object with lower luminosity has a higher mass than the object with greater luminosity. The ``dark'' object must have a mass of at least $1.4$~M$_\odot$, ensuring that it can only be a BH, NS, or a star mimicking the mass of a BH or NS} (see Section \ref{subsec:modelling}.)
  \item {{\bf Criteria 2:} The dark object can be either more or less massive than the bright object, but must still have a mass of at least $1.4$~M$_\odot$.}
\end{itemize}

\begin{table}
  \begin{center}
    \caption{Binary pairings which constitute false positives and true Unicorns.  Acronyms refer to:  BH=Black Hole, MS=Main Sequence Star, NHS=Naked Helium Star, NS=Neutron Star, RGB=Red Giant.}
    \label{tab:table3}
    \begin{tabular}{c|c}
        False Positive & True Unicorn\\
        \hline
        MS-MS & NS-MS\\
        RGB-MS & NS-RGB\\
        RGB-RGB & NS-NHS\\
        NHS-MS & BH-MS\\
        NHS-RGB & BH-RGB\\
         & BH-NHS\\
    \end{tabular}
  \end{center}
\end{table} 

For binaries that meet the false positive definition under Criteria 1, we can expect that a significant degree of stellar evolution has taken place. Unless the more massive companion is a main sequence star experiencing high mass loss or mass transfer, it is likely to be in its post-main sequence stage since it is less luminous. We also analyze our population under the Criteria 2 definition to ensure that we are not under-counting the numbers of true Unicorns and false positives, as binaries may contain a dim star or compact object which is less massive than its bright companion. An example of how these criteria are used to define an object at different times is discussed in Section \ref{subsec:Evolution}. Together, these two criteria set a lower and higher limit on the probability of potential false positives within a population. Table \ref{tab:table3} summarizes the binary types which we define as false positives and true Unicorns.

We also examine the numbers of true Unicorns and false positives under five different luminosity ratios:
\begin{itemize}
  \item {{\bf L$_{\mathrm{Dark}}$<L$_{\mathrm{Bright}}$:} The dark object is less luminous than the bright object by any amount.}
  \item {{\bf L$_{\mathrm{Dark}}$<1.5 L$_{\mathrm{Bright}}$:} Dark object is at least 1.5 times less luminous.}
  \item {{\bf L$_{\mathrm{Dark}}$<3 L$_{\mathrm{Bright}}$:} Dark object is at least 3 times less luminous.}
  \item {{\bf L$_{\mathrm{Dark}}$<5 L$_{\mathrm{Bright}}$:} Dark object is at least 5 times less luminous.}
  \item {{\bf L$_{\mathrm{Dark}}$<10 L$_{\mathrm{Bright}}$:} Dark object is at least 10 times less luminous.}
\end{itemize}

SED analysis of the false positive V723 Mon shows that the subgiant companion has $\sim$2/3 the luminosity of the giant primary star \citep{ElBadry2022BinaryZoo}. We can reasonably assume that a binary with this luminosity ratio or greater would require spectral disentangling to identify the ``dark'' companion. By considering different luminosity ratios, we place higher and lower limits on the probability of false positives based on how much less luminous we define the ``dark'' companion to be.

Our population of interest is the number of true Unicorns and false positives found within each time interval, with the count weighted to reflect what an observer would instantaneously find at that time (see Section \ref{subsec:modelling}). We define a "false positive probability" as the percentage of binaries which might appear to contain a compact object candidate but are in fact made up of two stars. We define a "true Unicorn probability" as the percentage of binaries which might appear to be contain a candidate and do actually contain a BH or NS.

\section{Results and Discussion}\label{sec:discussion}

\subsection{Kroupa Initial Mass Function}\label{subsec:KroupaIMF}

\begin{figure*}
     \centering
     \includegraphics[width=\linewidth]{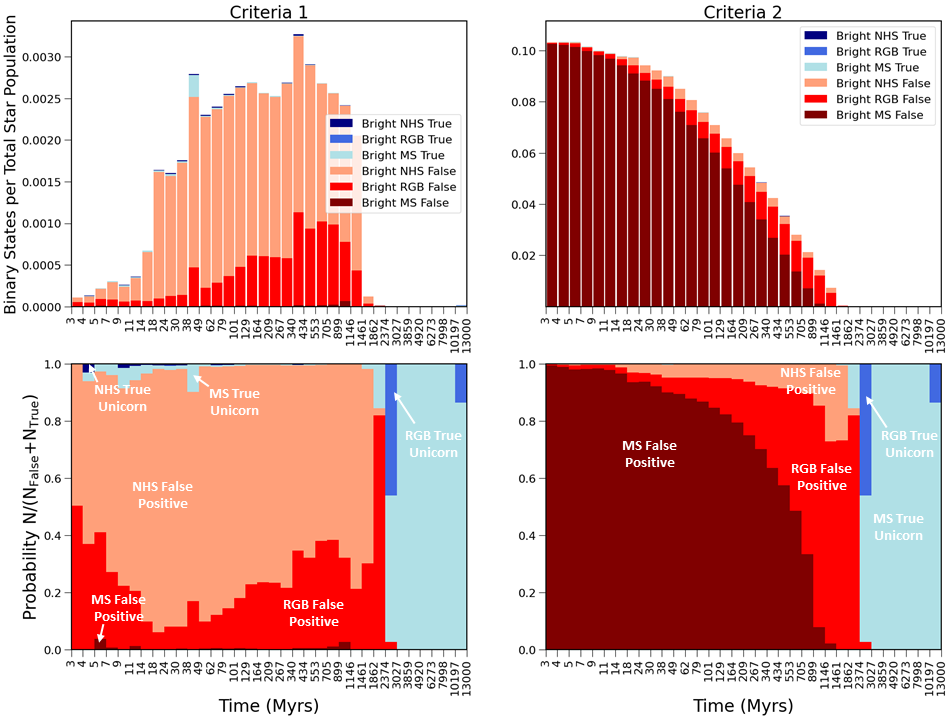}
     \caption{Results of stellar binary evolution modelling under a Kroupa IMF, focusing on the probabilities of false positives (red shades) and true Unicorns (blue shades) within the population. \textit{Top left:} shows the rate at which these objects appear at different time intervals as a percentage of the overall population of $5\times10^5$ binaries. This plot utilizes Criteria 1, wherein we require the dark object to be more massive than the bright object, and assumes L$_{\mathrm{Dark}}$<L$_{\mathrm{Bright}}$ by any amount. \textit{Bottom left:} provides the false positive and true Unicorn probabilities within our population of interest under Criteria 1. This gives the likelihood that any given binary with a bright star and a more massive dark object greater than $1.4$~M$_\odot$ forms a true Unicorn. \textit{Top right:} shows the rate at which these objects appear as a percentage of overall population under Criteria 2, which is a looser criteria wherein we do not require the dark object to be more massive than the bright object (note the change in y-axis scale from Top left plot). \textit{Bottom right:} shows the true Unicorn and false positive rates under Criteria 2. Acronyms refer to:  BH=Black Hole, MS=Main Sequence Star, NHS=Naked Helium Star, NS=Neutron Star, RGB=Red Giant.}
     \label{tab:Figure1}
 \end{figure*}

Figure \ref{tab:Figure1} shows the results of our stellar evolution models for the population formed under a Kroupa IMF \citep{Kroupa1993, Salpeter1955}, giving the probability that objects of interest are either a true Unicorn or false positive during different time intervals. Under Criteria 1, in which we assume the dark object to be greater in mass than the bright object (Figure \ref{tab:Figure1}, bottom left), we observe that binaries begin evolving into true Unicorns as early as 5 Myrs after cluster formation. The total false positive probability is high at early ages, remaining >90\% until 2.4 Gyrs. By 3.8 Gyrs, the false positive probability falls to near zero, which could in part be explained by a deficit of massive stars at late ages. Of the $5\times10^{5}$ binaries in our population, 385 ($0.08\pm0.01$\%) form a true Unicorn at some point in their evolution, while 34,019 ($6.80\pm0.04$\%) form a false positive.

Under Criteria 1, we observe that $\sim$77\% of the time a false positive is identified within various time intervals, the object is a NHS/MS binary with the NHS being the more luminous star. Such binaries have been proposed as explanations for several BH candidates, including LB-1 \citep{Liu2019, Irrgang2020} and NGC 1850 \citep{El-Badry2022AnotherImposter}. The next most common false positives are RGB/RGB binaries ($\sim$13\%), and RGB/MS binaries with the RGB the more luminous star ($\sim$9\%).

\begin{table}
  \begin{center}
    \caption{Binaries within our synthetic population similar to V723 Mon, as evolved under a Kroupa IMF \citep{Kroupa1993}, and based on the mass and temperature estimates in \citep{ElBadry2022BinaryZoo}. This shows that our simulations produce objects similar to a known false positive, although they do not reproduce V723 Mon exactly.}
    \label{tab:table4}
    \begin{tabular}{c|c|c|c|c}
        System & $T_{\rm eff, giant}$ & $M_{\rm giant}$ & $T_{\rm eff, 2}$ & $M_{2}$\\
         & (kK) & (M$_{\odot})$ & (kK) & (M$_{\odot})$\\
        \hline
        V723 Mon & 3.8$\pm$0.1 & 0.44$\pm$0.06 & 5.8$\pm$0.2 & 2.8$\pm$0.3\\
        \hline
        Sample 115677 & 4.2 & 0.41 & 5.3 & 3.3\\
        Sample 477609 & 4.2 & 0.41 & 5.1 & 3.2\\
    \end{tabular}
  \end{center}
\end{table}

Binaries similar to V723 Mon, with an RGB undergoing severe stripping (mass <0.5 M$_{\odot}$) and a less luminous subgiant companion (T$_{\rm eff}$ between 5000-6000 K), make up <1\% of false positive detections. While our simulations do not reproduce an exact replica of V723 Mon, they produce binaries with masses and temperatures similar to those calculated in \citet{ElBadry2022BinaryZoo}, confirming {\tt COSMIC}'s ability to reasonably reproduce known false positives. Table \ref{tab:table4} compares two of our synthetic binaries to V723 Mon.

 \begin{figure*}
     \centering
     \includegraphics[width=0.99\linewidth]{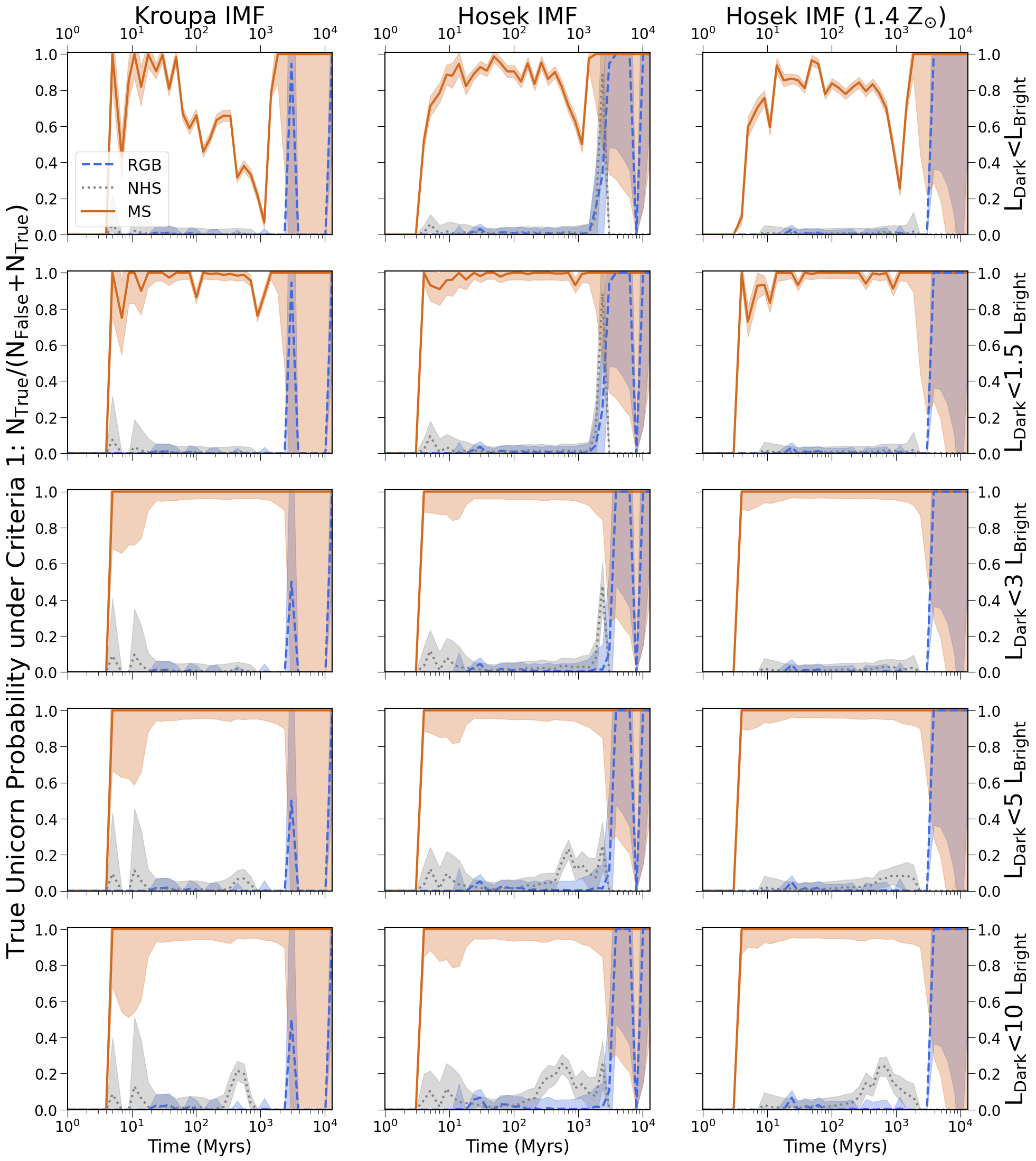}
     \caption{Results of stellar binary evolution modelling, providing the likelihood that an object of interest will be a true Unicorn at different times. True Unicorns are defined under Criteria 1, wherein we require the dark object to be more massive than the bright object. The MS true Unicorn Probability is in orange, RGB true Unicorn probability is in blue, and NHS true Unicorn probability is in gray. The colored regions above and below each line indicate the uncertainty. \textit{Column 1}: binaries are formed under a Kroupa IMF. \textit{Column 2}: binaries formed under a Hosek IMF. \textit{Column 3}: binaries formed under a Hosek IMF with metallicity of 1.4 Z$\odot$. \textit{Row 1}: the dark companion in the binary can be less luminous than the bright star by any amount (L$_{\mathrm{Dark}}$<L$_{\mathrm{Bright}}$). \textit{Rows 2, 3, 4, and 5}: L$_{\mathrm{Dark}}$<1.5 L$_{\mathrm{Bright}}$, L$_{\mathrm{Dark}}$<3 L$_{\mathrm{Bright}}$, L$_{\mathrm{Dark}}$<5 L$_{\mathrm{Bright}}$, L$_{\mathrm{Dark}}$<10 L$_{\mathrm{Bright}}$, respectively.}
     \label{tab:Figure2}
 \end{figure*}

 Figure \ref{tab:Figure2} shows the true Unicorn probabilities as defined under Criteria 1 for MS, RGB, and NHS, using each of the different IMFs and luminosity ratios in this study. The colored region above and below each line indicates the calculated level of uncertainty. Under the most basic luminosity ratio of L$_{\mathrm{Dark}}$<L$_{\mathrm{Bright}}$ (Figure \ref{tab:Figure2}, upper left), the likelihood of observing an RGB true Unicorn remains low for most ages, rising briefly to 94.6\% at 3.0 Gyrs (uncertainty is $\sim$100\% due to small number statistics). NHS true Unicorns are also unlikely, with probability <2\% for all time intervals. For MS stars, the true Unicorn probability rises to $85.7\pm11.1$\% at 9 Myrs, and remains between 80-100\% ($\pm4-13\%$) from 11-49 Myrs. Probability then falls gradually from $67.0\pm3.5$\% at 62 Myrs to $6.6\pm3.1$\% at 1.1 Gyrs, then rises to $78.7\pm3.4$\% at 1.5 Gyrs and $100\pm16.1$\% at 1.9 Gyrs. After this age, probability remains 100\%, but uncertainty is also $\sim$100\% due to small number statistics. At 13 Gyrs, uncertainties decrease slightly for RGB ($\pm$52.0\%) and MS stars ($\pm$73.2\%), due in part to the larger size of this final logarithmic time bin.

 As we consider progressively stricter luminosity ratios (Figure \ref{tab:Figure2}, left column), changing our definition of how the ``dark'' object is defined in each binary, the number of true Unicorns decreases only slightly with each change. This is due to  hot, recently-formed neutron stars (<100 Myrs in age) close to the luminosity of their small stellar companions being removed as the luminosity ratio changes. The number of binaries forming a true Unicorn at some point in their evolution dips to 382 for L$_{\mathrm{Dark}}$<3 L$_{\mathrm{Bright}}$ and 380 for L$_{\mathrm{Dark}}$<10 L$_{\mathrm{Bright}}$. The number of false positives decreases more rapidly, driving the overall change in true Unicorn probability with each different luminosity ratio. The total number of binaries forming a false positive at some point in their evolution falls to 27,516 ($5.50\pm0.03$\%) for L$_{\mathrm{Dark}}$<3 L$_{\mathrm{Bright}}$, and 24,103 ($4.82\pm0.03$\%) for L$_{\mathrm{Dark}}$<10 L$_{\mathrm{Bright}}$.

 Under different luminosity ratios, MS true Unicorns become more likely. Defining the ``dark'' companion under L$_{\mathrm{Dark}}$<1.5 L$_{\mathrm{Bright}}$, MS true Unicorn probability rises to $100.0\pm24.2$\% at 5 Myrs, dips to $75.0\pm20.1$\% at 7 Myrs, then rises back to $\sim$100\% ($\pm3-17$\%) from 9-79 Myrs, 129-705 Myrs, and 1.5-2.4 Gyrs. Under L$_{\mathrm{Dark}}$<3 L$_{\mathrm{Bright}}$, the MS true Unicorn probability is $100.0\pm31.7$\% at 5 Myrs, and remains at this level ($\pm$4-34\%) until 3.0 Gyrs, where small-number statistics drive uncertainties upward. For binaries with a bright RGB, the true Unicorn probability changes little with different luminosity ratios. For binaries with a bright NHS, we see only one area where probability increases with changing luminosity ratios, from 434-553 Myrs.

\begin{figure*}
     \centering
     \includegraphics[width=0.99\linewidth]{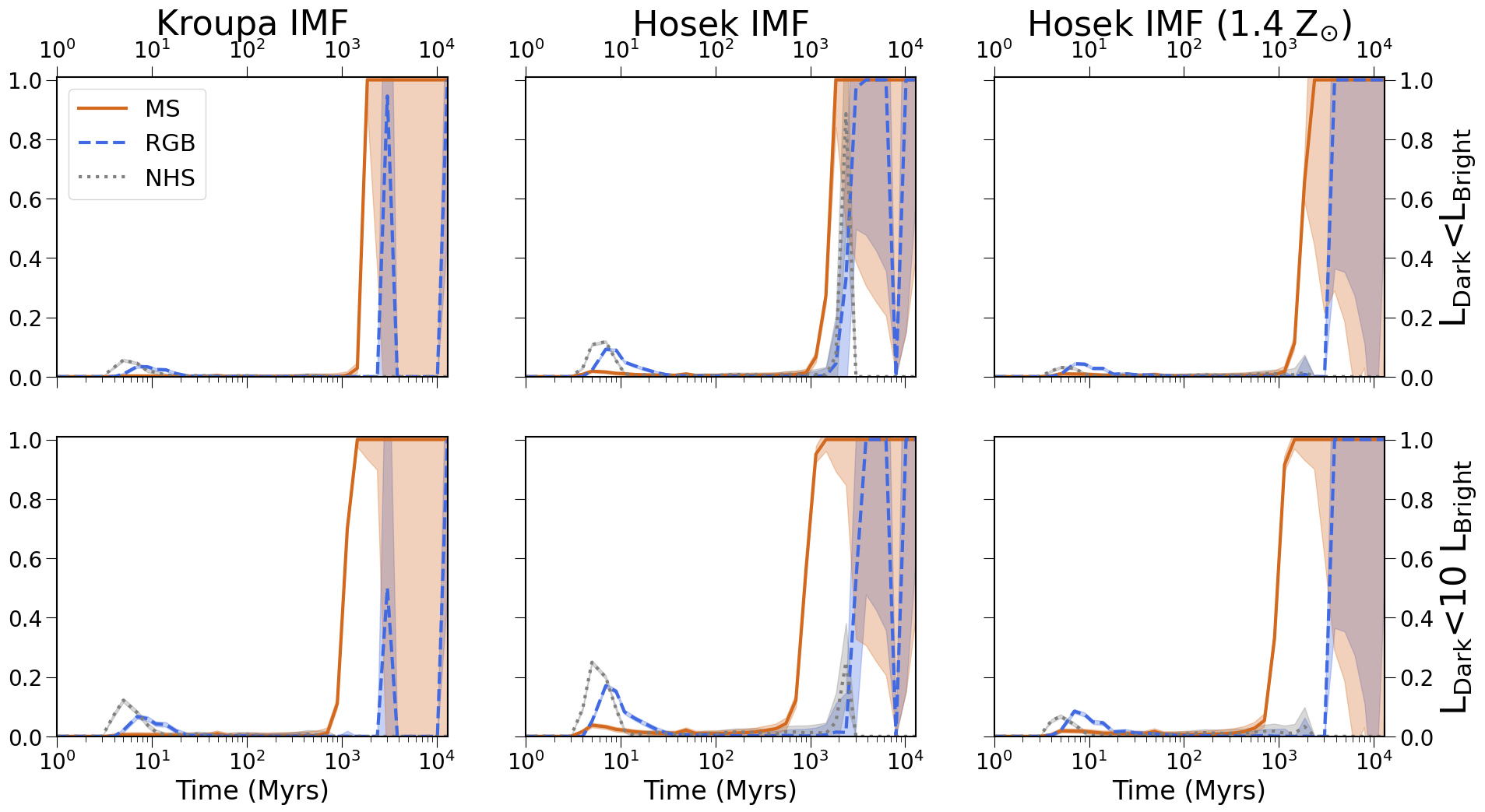}
     \caption{Results of stellar binary evolution modelling, providing the likelihood that an object of interest will be a true Unicorn at different times. The MS true Unicorn Probability is in orange, RGB true Unicorn probability is in blue, and NHS true Unicorn probability is in gray. The colored regions above and below each line indicate the uncertainty. True Unicorns are defined under Criteria 2, wherein we allow the dark object to be either more or less massive than the bright star. \textit{Column 1:} binaries are formed under a Kroupa IMF. \textit{Column 2:} binaries formed under a Hosek IMF. \textit{Column 3:} binaries formed under a Hosek IMF with metallicity of 1.4 Z$\odot$. \textit{Row 1:} the dark companion in the binary can be less luminous than the bright star by any amount (L$_{\mathrm{Dark}}$<L$_{\mathrm{Bright}}$). \textit{Row 2:} the dark companion has < 10 times the luminosity of the bright star (L$_{\mathrm{Dark}}$<10 L$_{\mathrm{Bright}}$). Additional luminosity ratios are not shown here, as they are qualitatively similar to those shown.}
     \label{tab:Figure3}
 \end{figure*}
 
Under Criteria 2, in which we assume the dark object is more massive than $1.4$~M$_\odot$ but may be more or less massive than the bright star, we observe a higher number of true Unicorns and a much higher number of false positives compared to Criteria 1. 700 binaries ($0.14\pm0.01$\%) form true Unicorns at some point in their evolution, while 65,644 ($13.1\pm0.1$\%) form false positives. Both categories are roughly doubled from Criteria 1. 

In Figure \ref{tab:Figure1} (bottom right), we see that the total false positive probability under Criteria 2 remains above 99\% until 2.4 Gyrs, then decreases to near-zero at 3.0 Gyrs. $\sim$89\% of the time a false positive is identified, the object is a MS/MS binary. This is followed by RGB/MS binaries, with the RGB the more luminous star ($\sim$7\%), and NHS/MS binaries, with the NHS the more luminous star ($\sim$3\%.)

Figure \ref{tab:Figure3} shows the true Unicorn probabilities as defined under Criteria 2 for MS, RGB, and NHS, using each of the different IMFs and two of the luminosity ratios in this study (the others are qualitatively similar, and are not shown). As before, the colored region above and below each line indicates the calculated uncertainty. The RGB true Unicorn probability (Figure \ref{tab:Figure3}, upper left) remains low for most ages, climbing to $\sim$95\% at 3.0 Gyrs, where uncertainty is again $\sim$100\% due to small number statistics. At 13 Gyrs, uncertainty dips as it did under Criteria 1. NHS true Unicorn probability also remains low, with a peak of $\sim5.0\pm0.5$\% from 4-7 Myrs.  MS true Unicorn probability is much lower than under Criteria 1, remaining below 1\% until 1.5 Gyrs. It then climbs to $100.0\pm8.2$\% at 1.9 Gyrs, remaining high through 2.3 Gyrs ($100.0\pm63.0$\%). At 3.0 Gyrs uncertainty again reaches $\sim$100\%. We observe that ``dark'' objects are considerably more likely to be a BH or NS if they meet the Criteria 1 definition of being more massive than the bright MS star. While this makes intuitive sense, the level of difference between probabilities under Criteria 1 and Criteria 2 is striking.

 As we compare different luminosity ratios under Criteria 2 (Figure \ref{tab:Figure3}, left column), we observe very small increases in RGB and NHS true Unicorn probabilities.  MS true Unicorn probability under L$_{\mathrm{Dark}}$<10 L$_{\mathrm{Bright}}$ increases by <1\% compared to L$_{\mathrm{Dark}}$<L$_{\mathrm{Bright}}$ until 899 Myrs, when it reaches $11.0\pm1.8$\%, then $70.0\pm1.8$\% at 1.1 Gyrs and 100\% from 1.5 to 3.0 Gyrs ($\pm$2-10\%). Although the MS true Unicorn probability is low under Criteria 2, there is a window of time from $\sim$1-3 Gyrs where true Unicorns are more likely if the ``dark'' object is confirmed to be significantly less luminous than the bright star.

\subsection{Hosek Initial Mass Function}\label{subsec:HosekIMF}

Within our population produced under the IMF described by \citet{Hosek2019}, the total number of both true Unicorns and false positives is higher than under a Kroupa IMF due to the greater fraction of massive stars. Under Criteria 1, we find that 776 objects ($0.16\pm0.01$\%) form true Unicorns and 47,781 ($9.56\pm0.04$\%) form false positives at some point in their evolution. Under Criteria 2, we find that 4102 objects ($0.82\pm0.01$\%) form true Unicorns and 90,476 ($17.1\pm0.1$\%) form false positives. A plot of the total true Unicorn and false positive probabilities within each time bin can be found in Appendix \ref{sec:appendixB}. Under Criteria 1 (Figure \ref{tab:FigureB1}, bottom left), the total false positive probability remains $\ge$89\% until 1.9 Gyrs, where it declines to 66\%. By 3.0 Gyrs the false positive rate drops to near-zero (small-number statistics remain an issue at this age, although less so than under a Kroupa IMF). Under Criteria 2 (Figure \ref{tab:FigureB1}, bottom right), the total false positive probability remains $\ge$98\% until 1.9 Gyrs, where it declines to 83\%. By 3.0 Gyrs it drops to near-zero.

Under Criteria 1 and the basic luminosity ratio of L$_{\mathrm{Dark}}$ < L$_{\mathrm{Bright}}$, the RGB true Unicorn probability (Figure \ref{tab:Figure2}, upper middle) is only $\sim$1\% higher than under a Kroupa IMF until 1.9 Gyrs, where it rises to $14.3\pm23.2$\%, then to $100.0\pm52.2$\% at 3.9 Gyrs. NHS true Unicorn probability is also $\sim$1\% higher than under a Kroupa IMF, rising to $88.6\pm38.4$\% at 2.4 Gyrs, and declining to near-zero at 3.0 Gyrs. At early ages, MS true Unicorn probability does not follow an obvious trend compared to the Kroupa IMF, becoming 50\% more likely at 4 Myrs, 30\% less likely at 5 Myrs, 35\% more likely at 7 Myrs, and so on. A trend arises in the 62 Myr to 1.5 Gyr timeframe, where MS true Unicorn probability is 30-60\% higher than a Kroupa IMF, reaching $97.4\pm5.6$\% at 1.5 Gyrs and $100.0\pm27.6$\% at 1.9 Gyrs. While uncertainties beyond this age are still high ($\sim$50-80\%), they are lower than under a Kroupa IMF.

Under Criteria 2, the RGB true Unicorn probability (Figure \ref{tab:Figure3}, upper middle) is a few percent higher than under a Kroupa IMF from 5-24 Myrs. At 3.0 Gyrs, it rises to $97.2\pm47.5$\%, then averages $\sim100\pm56$\% afterward. NHS true Unicorn probability is also a few percent higher from 4-9 Myrs, peaking at 2.4 Gyrs ($88.6\pm37.1$\%) before falling to near-zero. MS true Unicorn probability increases by $\sim$1\% from 5-9 Myrs, then peaks at 1.5 Gyrs ($27.2\pm2.8$\%). As before, probability is $\sim$100\% from 1.9 Gyrs on, with lower uncertainties than under a Kroupa IMF (averaging $\pm$70\%).

Comparing different luminosity ratios, as we advance from L$_{\mathrm{Dark}}$<L$_{\mathrm{Bright}}$ to L$_{\mathrm{Dark}}$<10 L$_{\mathrm{Bright}}$ under Criteria 1 (Figure \ref{tab:Figure2}, middle column), we observe increases of a few percent in the RGB true Unicorn probability at early times, with significant decreases at middle ages. Under Criteria 2 (Figure \ref{tab:Figure3}, middle column), we observe a 1-8\% increase from 5-30 Myrs and greater decreases from 1.9-3.0 Gyrs. The NHS true Unicorn probability under Criteria 1 rises by a few percent as we advance from L$_{\mathrm{Dark}}$<L$_{\mathrm{Bright}}$ to L$_{\mathrm{Dark}}$<10 L$_{\mathrm{Bright}}$ (5-25\% from 209 Myrs-1.5 Gyrs). Under Criteria 2, NHS true Unicorn probability increases by 4-14\% during the 4-9 Myr timeframe, but decreases significantly at 2.4 Gyrs. For MS stars under L$_{\mathrm{Dark}}$<10 L$_{\mathrm{Bright}}$, the MS true Unicorn probabililty is $\sim$100\% for early ages under Criteria 1, with uncertainties of $\le$20\%. After 3.0 Gyrs, the average uncertainty is $\pm$77\%. Under Criteria 2, we again find a window where probabililty increases but uncertainties are still low. At 899 Myrs, probability reaches $55.8\pm2.6$\%, then $\sim$100\% from 1.1 to 3.0 Gyrs ($\pm$8\% on average) . During this time period, we observe a high MS true Unicorn probability when the ``dark'' object is much less luminous than the bright star.

We observe that in general, populations formed under a Hosek IMF are more likely to produce MS stars orbiting a compact object than a Kroupa IMF, particularly when the ``dark'' object is more massive than the bright star. This is a reasonable expectation, as a top-heavy IMF produces a greater number of massive stars, and therefore a greater number of compact object companions. At its current age, $3.7\pm0.2$~Myrs \citep{Hosek2019,Gallego2021}, the Arches Cluster should already be forming a small population of these binaries, with a MS true Unicorn probability of $52.3\pm4.0$\% compared to $\sim$0\% under a Kroupa IMF. Top-heavy-IMF star clusters represent a potentially productive environment to search for non-interacting compact object binaries. However, this is only taking into account the IMF and ignoring other differences between stellar populations.

\subsection{Hosek Initial Mass Function with Super-Solar Metallicity}\label{subsec:ArchesHighMetallicity}

For the stellar populations discussed in Sections \ref{subsec:KroupaIMF} and \ref{subsec:HosekIMF}, we have used solar metallicity for all stars \citep{Asplund2009}. However, observation show that the Arches Cluster, upon which the Hosek IMF is based, may exhibit a super-solar metallicity of 1.3-1.4 Z$_\odot$ \citep{Martins2008, Sabhahit2022}. To more realistically model the Arches Cluster, which we use to represent young star clusters in general, we evolve an additional population of $5\times10^{5}$ binaries in {\tt COSMIC} using the \citet{Hosek2019} IMF and a higher metallicity value of  $Z=0.01876$.
 
Under Criteria 1 (Figure \ref{tab:Figure2}, upper right), we observe decreases of $\sim$1\% in the RGB true Unicorn probability compared to the 1 Z$_\odot$ Hosek population. At 1.9 Gyrs, probability drops to zero. At 3.9 Gyrs, probability rises to $100.0\pm63.6$\%. The NHS true Unicorn probability also sees a small decrease, then drops to zero at 2.4 Gyrs. We observe a decrease in the MS true Unicorn probability of 40\% at 4 Myrs, then a smaller decrease of $\sim$3-10\% from 5-705 Myrs and 15-25\% from 899 Myrs-1.5 Gyrs. At 1.9 Gyrs, probability reaches $100.0\pm11.7$\%, after which it remains $\sim$100\% ($\pm$93\% on average).

Under Criteria 2 (Figure \ref{tab:Figure3}, upper right), we observe decreases of a few percent in the RGB true Unicorn probability compared to the 1 Z$_\odot$ population, with a drop to zero at 2.4-3.0 Gyrs. Probability remains $\sim$100\% from 3.9 Gyrs onward ($\pm$77\% on average). The NHS true Unicorn probability is low for most ages ($\le$1\%). We observe small decreases in MS true Unicorn probability compared to the 1 Z$_\odot$ population, with larger decreases of $\sim$15\% at 1.5 Gyrs and $\sim$35\% at 1.9 Gyrs. From 2.4 Gyrs onward, probability remains $\sim$100\% ($\pm$93\% on average).

These decreases in true Unicorn probability agree with expectation. Under a metallicity slightly higher than solar value, BH formation will slow as massive stars experience high mass loss, causing them to collapse into neutron stars \citep{Heger2003}. For stellar populations formed under a super-solar metallicity, any true Unicorn is more likely to contain a NS than a BH. We observe an $\sim$11\% increase in the number of NS under 1.4 Z$_{\odot}$.

The effects of a change in luminosity ratio have been discussed in Sections \ref{subsec:KroupaIMF} and \ref{subsec:HosekIMF}, and are qualitatively similar for this population (Figures \ref{tab:Figure2} and \ref{tab:Figure3}, right columns). For binaries defined under L$_{\mathrm{Dark}}$<10 L$_{\mathrm{Bright}}$, we still observe a significant rise in MS true Unicorn probability from 1.1-3.9 Gyrs, with relatively low uncertainties ($\pm$14\% on average). For MS stars of this age, false positives are unlikely to be found among ``dark'' objects significantly less luminous than their bright companions.

While a super-solar metallicity does decrease the likelihood of the Arches Cluster forming true Unicorns at its current age, $3.7\pm0.2$~Myrs \citep{Hosek2019,Gallego2021}, we still find a $9.6\pm3.8$\% MS true Unicorn probability at 4 Myrs under Criteria 1 (requiring that the companion is more massive than the MS star). This increases to $100.0\pm6.5$\% if the companion is confirmed to be at least 1.5 times less luminous than the bright star. The Arches Cluster and any similar young clusters with top-heavy IMFs should be relatively productive environments for locating true Unicorns, even under higher metallicity.
 
\subsection{Orbital Characteristics}\label{subsec:OrbitalCharacteristics}

\begin{figure}
     \centering
     \includegraphics[width=\linewidth]{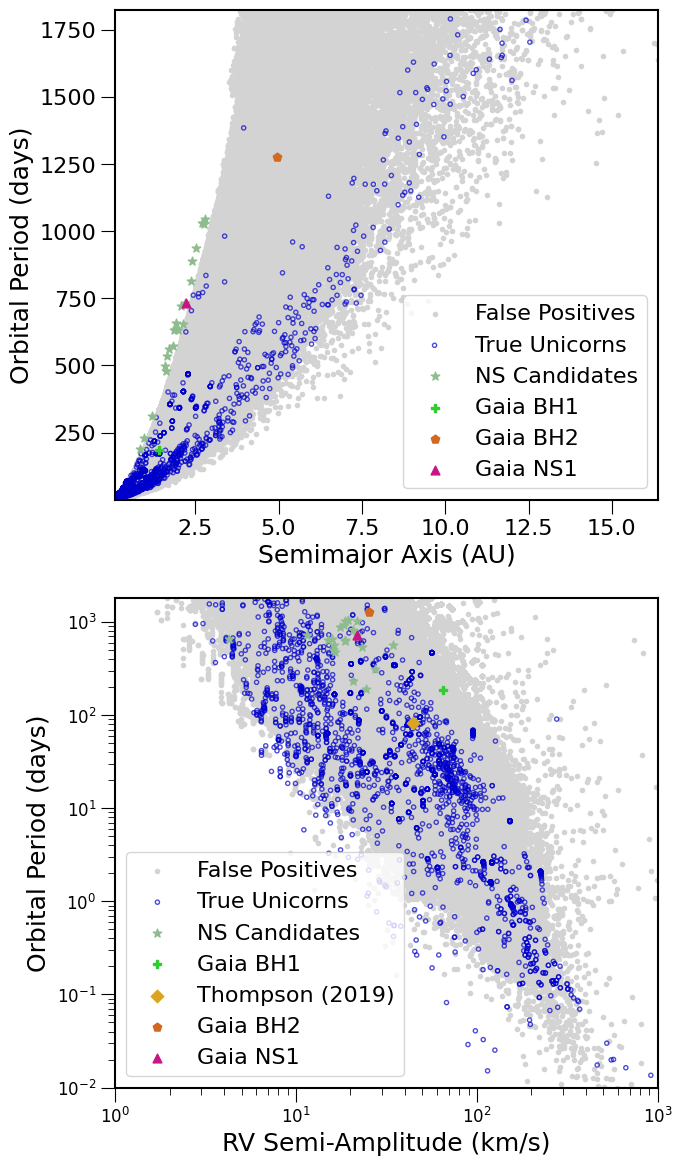}
     \caption{Orbital characteristics of stars in the synthetic binary population formed under a Kroupa IMF. \textit{Top plot:} Orbital period in days vs. semi-major axis in astronomical units (AU) for all objects which we identify as true Unicorns or false positives under Criteria 2. Colored points indicate known objects with compact object companions, described in Section \ref{sec:introduction}. \textit{Bottom plot:} Orbital period in days vs. radial velocity semi-amplitude in km/s for false positives and true Unicorns.}
     \label{tab:Figure4}
 \end{figure}
 
We seek to examine whether true Unicorns might have unique orbital parameters distinct from false positives. Figure \ref{tab:Figure4} (top plot) shows the orbital period vs. semi-major axis at various times (provided by {\tt COSMIC}) of the binaries evolved under a Kroupa IMF which we identify as true Unicorns and false positives in our model. Figure \ref{tab:Figure4} (bottom plot) shows the orbital period provided by {\tt COSMIC} vs. the radial velocity semi-amplitude (as defined in Section \ref{subsec:measuringradialvelocities}) of the more luminous object in each pair. These plots use the definition of Criteria 2, where the dark object can be either more or less massive than the bright star, and L$_{\mathrm{Dark}}$<L$_{\mathrm{Bright}}$, where the dark star can be less luminous than the bright star by any amount. At more restrictive luminosity ratios, the number of false positives decreases overall, but the distributions remain similar. Also included are the orbital characteristics of known objects (see Section \ref{sec:introduction}), including 2MASS J05215658+435922 \citep{Thompson2019}, Gaia BH1 \citep{El-Badry2023}, Gaia BH2 \citep{El-Badry2023RG}, Gaia NS1 \citep{ElBadry2024NS1}, and the NS candidates discovered in \citet{ElBadry2024NScand}.

Comparing orbital period vs. semimajor axis (Figure \ref{tab:Figure4}, top plot), we observe that most true Unicorns are clustered in a low-period regime. True Unicorns form an orbital period >400 days only $11.8\pm1.0$\% of the time, and >100 days only $26.4\pm1.4$\% of the time. A review of \textit{Gaia} data shows that compact object companions should be found more often at shorter periods \citep{El-Badry2023}, and our models agree with this.  Under Criteria 1, only $13.2\pm1.8$\% of true Unicorns have a period >100 days, indicating that periods are even shorter when the dark object is more massive than the bright star.

We also observe that for most true Unicorns with a period equal to that of a false positive, the true Unicorn has a wider orbital separation. This again matches expectation. NS and BH masses tend to be higher than the average star, and Kepler's third law stipulates that more massive objects in a binary will produce a wider orbit at a given period \citep{El-Badry2023}. Age-related rotational slowing \citep{Barnes2003} can also play a part in widening orbits, as can supernova kicks. In Figure \ref{tab:Figure4} (top plot), several of the discovered BH and NS binaries lie outside the true Unicorn parameter space of the plot, however this is simply due to the high starting mass of our synthetic population (each of our binaries contains a primary star with initial mass $\ge$2 M$_{\odot}$). If we were to plot the stars within each object's local population, we would see their orbits as being wider than the stars around them.

Comparing orbital period vs. RV semi-amplitude (Figure \ref{tab:Figure4}, bottom plot), we find that true Unicorns do not appear to follow a particular trend compared to false positives. However, true Unicorns do become less common at orbits <10$^{-1}$ days (since small periods are likely dominated by x-ray binaries), and at radial velocities >400 km/s. Many of the actual objects discovered fit within the true Unicorn parameter space.

Under a Hosek IMF (both 1 Z$_\odot$ and 1.4 Z$_\odot$ metallicities), we observe qualitatively similar period vs. semi-major axis and period vs. semi-amplitude distributions. The only significant difference is a higher overall number of true Unicorns and false positives.

\subsection{Example Evolution}\label{subsec:Evolution}

\begin{figure}

     \centering
     \includegraphics[width=0.99\linewidth]{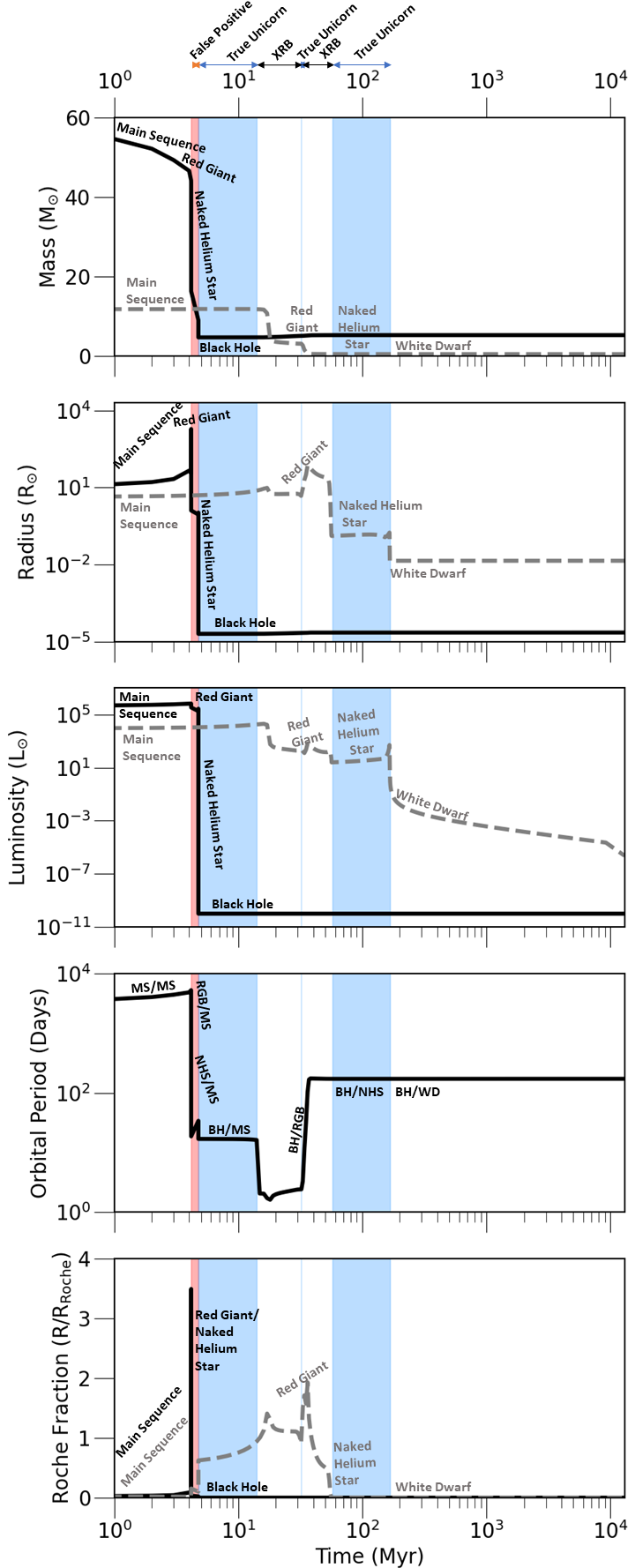}
     \caption{Representative example of the evolution of a stellar binary pair from our synthetic population. The object is sampled at 1 Myr timesteps and at any key changes in stellar evolution. Mass (M$_\odot$), radius (R$_\odot$), visible luminosity (L$_\odot$), period (days), and Roche lobe filling fraction (R/R$_{\rm Roche}$) are given for both stars. Initial mass of the primary star is 56.5 M$_{\odot}$, and initial mass of the secondary is 11.8 M$_{\odot}$. Initial orbital period is 3516 days. This plot demonstrates how a binary can change definition over its lifetime.}
     \label{tab:Figure5}
 \end{figure}

Figure \ref{tab:Figure5} provides an example of the stellar evolution of a single binary under our model. The purpose of this plot is to illustrate how the categories we define for each object of interest (true Unicorn, false positive, x-ray binary) are defined at a given time, and do not apply to the entire lifetime of the binary. The evolution of a binary pair may carry it through many stages over its lifetime. Masses, radii, luminosities, and Roche lobe filling fractions are given for both stars, as well as the orbital period. This binary is somewhat atypical within our populations given its high initial mass, however it illustrates the various states of our model well.

{\tt COSMIC} tracks binary evolution in two ways: 1) by sampling the binary characteristics each time a key evolutionary change occurs; and 2) by sampling the binary parameters at a series of user-defined time steps, which we have set to 1 Myr for this particular binary. As before, we combine the time step output with the output from evolutionary changes to ensure that no significant developments were missed.

The binary begins as a MS/MS pair, then evolves to a RGB/MS pair at 4.145 Myrs, which we define as a false positive under Criteria 2 (a bright RGB with a darker, less massive MS companion). The pair then evolves into a NHS/MS binary at 4.151 Myrs, which meets the definition of a false positive under Criteria 1 (a bright MS star and a darker, more massive NHS companion). At 4.719 Myrs, the NHS has lost mass, and the object returns to a false positive definition under Criteria 2 (a bright MS star with a darker, less massive NHS.) At 4.748 Myrs, the pair evolves into a true Unicorn as defined under Criteria 2 (a BH/MS pair with the compact object less massive than the bright star).  

At 14 Myrs, the MS star (with a current mass of 11.8 M$_{\odot}$) fills >90\% of its Roche lobe, and the BH forms an accretion disk with an x-ray luminosity >10$^{35}$ erg s$^{-1}$. This creates a high-mass x-ray binary (HMXB). The BH is now detectable through x-ray emission, and no longer meets the conditions of a non-interacting true Unicorn. As the BH accretes mass, the MS star shrinks to 4.2 M$_{\odot}$ at 18 Myrs, changing from a HMXB to an IMXB (intermediate-mass x-ray binary). At 31.9 Myrs, the bright star evolves onto the red giant branch, and accretion briefly slows. The binary becomes a true Unicorn under Criteria 1 (a bright RGB star and a more massive non-interacting BH).

At 32.1 Myrs, the RGB overflows its Roche lobe, and the binary becomes an IMXB once again. The orbital period also begins to expand. At 36 Myrs, the RGB's mass falls below 1 M$\odot$, and classification changes to a low mass x-ray binary (LMXB). At 56.3 Myrs, the bright RGB star has lost sufficient mass to form a NHS. The mass loss rate slows, and at 57 Myrs, the x-ray luminosity of the BH accretion disk drops to 3.8$\times$10$^{31}$ erg s$^{-1}$. The binary is no longer detectable through x-ray emission. It is again considered a true Unicorn under Criteria 1 (a NHS with a more massive, non-interacting BH companion).

At 165.9 Myrs, the NHS condenses into a white dwarf, and is no longer an object of interest for our model. 

\subsection{X-Ray Binary Evolution}\label{subsec:XRBEvolution}

The focus of our study is to identify the number of binary systems where object classification is uncertain and it is difficult to differentiate between a compact object and a massive, low-luminosity star. Since x-ray binaries provide direct evidence of compact objects through the presence of an accretion disk and detectable x-ray radiation, we have removed interacting x-ray binaries from our population within each time interval (see Section \ref{subsec:xraybinaries}). However, many true Unicorns do form x-ray binaries at some point during their evolution. In relatively short orbits, periods of mass transfer are likely to occur. In a typical star-forming galaxy, we expect that $\ge$10\% of compact objects form high-mass x-ray sources at least once during their lifetimes \citep{Mineo2012, Gilfanov2023}. Within our more limited population of true Unicorns (with their defined orbital and mass constraints), we seek to determine whether compact objects form x-ray binaries at a similar rate.

Within our population of $5\times10^{5}$ binaries formed under a Kroupa IMF, 1265 objects form either LMXBs, IMXBs, and/or HMXBs at some point in their evolution. Of these, 191 also evolve into true Unicorns under Criteria 1 (424 under Criteria 2's definition). Under a Hosek IMF, 8924 objects form x-ray binaries at some point in their evolution, 425 of which also evolve into true Unicorns under Criteria 1 (3238 under Criteria 2).

Comparing to the totals of true Unicorns in Sections \ref{subsec:KroupaIMF} and \ref{subsec:HosekIMF}, we observe that $49.6\pm3.6$\% of true Unicorns formed under a Kroupa IMF and defined under Criteria 1 evolve into x-ray binaries ($60.6\pm2.9$\% under Criteria 2). Of these, $\sim$2\% under Criteria 1 form HMXBs ($\sim$33\% under Criteria 2). Under a Hosek IMF, $54.8\pm2.7$\% of true Unicorns defined under Criteria 1 evolve into x-ray binaries ($78.9\pm1.4$\% under Criteria 2). Of these, $\sim$4\% under Criteria 1 form HMXBs ($\sim$69\% under Criteria 2).

As a subset of the overall compact object population, around half of true Unicorns under Criteria 1 (the compact object is more massive than the bright star) form XRBs at some point in their evolution, though the rate at which they form HMXBs is below average. Given that more massive objects in a binary will produce wider orbits \citep{ElBadry2024NScand}, it is reasonable to expect a lower level of accretion for these pairs. True Unicorns defined under Criteria 2 (the compact object is either more or less massive than the bright star) account for a higher number of x-ray emitting pairs, and are also much more likely to form HMXBs than typical black holes.

In Sections \ref{subsec:KroupaIMF}, \ref{subsec:HosekIMF}, and \ref{subsec:ArchesHighMetallicity}, we have shown that binaries with a more massive compact object and a less massive stellar companion are more likely to be observed than binaries with a less massive compact object and a more massive stellar companion. Since these more massive compact objects are also less likely to emit detectable x-rays, the importance of radial velocity measurements and astrometry in combination with spectroscopy is well demonstrated. Even under a top-heavy IMF, which favors the formation of massive stars and compact objects, $\sim$45\% of BH and NS binaries in relatively close orbits will never be detectable through x-ray emission during a 13 Gyr lifetime.
 
\subsection{Assumptions and Caveats}\label{subsec:Caveats}
 
 We model all stars as having formed from a single starburst event. Star clusters such as the Arches cluster likely experience only one epoch of formation, after which the remaining gas is cleared by stellar winds \citep{Krause2020}. Within a broader galactic population, however, additional stars do form at later times following supernova events, which will likely add to the number of both false positives and true Unicorns over time. An exploration of this effect is beyond the scope of this work. Nevertheless, as long as the age of a bright star in a binary can be independently determined, our models provide a reasonable probability that the star is orbiting a non-interacting compact object.

 We have also modelled our population as $5\times10^5$ binary pairs, disregarding the evolutionary impact of higher-order multiples like triples and quadruples.  Generally, stellar multiplicity fractions increase with stellar primary mass, driving an increase in the number of triple stars, quadruple stars, and beyond \citep{Raghavan2010}. Additional multiplicity may introduce complex dynamics on a physical star population, such as Kozai-Lidov mechanisms \citep{Lidov1962, Kozai1962, Naoz2016}, tidal interactions \citep[e.g.,][]{Naoz2014}, gravitational wave events \citep[e.g.,][]{Hoang2018}, and dynamically driven stellar mergers \citep[e.g.,][]{Rose2023}.

Our sample of true Unicorns includes a significant population of binaries experiencing ongoing mass transfer. Our observational characterization of an "interacting binary" is that it produces x-rays. While x-ray binaries are the product of ongoing mass accretion, not all accreting compact objects will emit detectable x-rays. Indeed, for the Unicorn, \citet{Jayasinghe2021Unicorn} proposed an accretion disc around the black hole, despite no detected x-rays. The disc was used to explain a relative dilution of blue wavelengths in the SED. While we have removed x-ray binaries from our data within each time interval, some binary pairs in our data may be interacting in other ways, some of which may produce radiation in other wavelengths.

Massive stars are also capable of producing x-ray emission through means other than accreting onto a compact object \citep{Rauw2015}, some of which can produce a high enough x-ray luminosity to rival x-ray binaries. These processes include hydro-dynamic shocks produced by instability in the stellar wind \citep[e.g.,][]{Feldmeier1997}, the
collision of wind-driven magnetically channeled gas \citep[e.g.,][]{Babel1997, ud-Doula2002}, and large-scale shocks from wind-wind interactions between large binary stars \citep{Stevens1992}. These processes are not modeled within our simulations, however they would provide another method through which massive false positive stars could be identified and differentiated from compact objects through their x-ray emission.

\section{Conclusions}\label{sec:conclusion}

To investigate the formation and evolution of compact objects, it is essential to locate them in non-interacting binaries, however this can be difficult to do observationally. In our models, we used {\tt COSMIC} to generate synthetic populations of binary stars formed from both a Kroupa IMF \citep{Kroupa1993} and a Hosek IMF \citep{Hosek2019}. To effectively study compact object formation, we set the initial mass of the primary star in each binary to $\ge$2 M$_{\odot}$.  We counted the numbers of true Unicorns, defined as tightly-orbiting binaries containing a compact object more massive than $1.4$~M$_\odot$, and false positives, defined as tightly-orbiting binaries with one bright star and one dark star whose mass and luminosity mimic a compact object and would likely require spectral disentangling to characterize the binary. We define our "false positive probability" as the percentage of binaries which might appear to contain a compact object candidate but are in fact made up of two stars. We define our "true Unicorn probability" as the percentage of binaries which might appear to be contain a candidate and do actually contain a BH or NS. While other studies have examined the number of compact objects that should exist within the Galaxy, this work examines how often stellar binaries can mimic these compact objects. With this study, we make a number of predictions:

\begin{itemize}
  \item {Within a population of massive stars, main sequence stars have a significantly higher true Unicorn probability compared to red giants or naked helium stars, particularly at young ages. Under a \citet{Kroupa1993} IMF, if we require the dark companion to be more massive than the bright star (which we call Criteria 1), the true Unicorn probability is between 80-100\% for stars aged 9-49 Myrs, falls from $67.0\pm3.5$\% at 62 Myrs to $6.6\pm3.1$\% at 1.1 Gyrs, then rises again to $78.7\pm3.4$\% at 1.5 Gyrs and $100\pm16.1$\% at 1.9 Gyrs. After this age, small number statistics make prediction less certain. In clusters formed under a top-heavy \citet{Hosek2019} IMF, the MS true Unicorn probability is generally higher, and in particular for stars aged 62 Myr to 1.5 Gyr (probability increases by 30-60\%). The false positive probability decreases to near-zero if we require the ``dark'' companion in each binary to be $\ge$3 times less luminous than the bright star. Therefore we can predict conditions under which a main sequence star with a non-interacting compact object companion is substantially more likely than a false positive: for a bright, young MS star with a less luminous companion (provided the companion has mass >1.4 M$_{\odot}$ and is more massive than the MS star, and the orbital period is $\le$5 years) if the companion is confirmed to be at least 3 times less luminous than the MS star, there is an extremely high probability of the companion being a non-interacting compact object. No false positives were found orbiting young MS stars below this luminosity cutoff. This applies to the populations formed under a \citet{Hosek2019} IMF as well.}
  \item {Under Criteria 2, in which we do not require the dark companion to be more massive than the bright star, the probability of locating a non-interacting compact object compared to a false positive is generally low. For MS stars, we see that only stars aged 1-2 Gyrs have a high true Unicorn probability (which expands to 900 Myrs-3 Gyrs if the companion is significantly less luminous). While compact objects can be less massive than their stellar companions, the probability of false positives is high for those objects.}
  \item {If we require the less luminous object in a binary to be more massive (Criteria 1), the most common type of false positives formed are NHS/MS binaries ($\sim$77\%), with the NHS (an evolved star severely stripped by mass loss, leaving the core exposed) being the more luminous object. If we do not require the less luminous object to be more massive (Criteria 2), the most common false positives are MS/MS binaries ($\sim$89\%).}
  \item {A super-solar metallicity will reduce the number of black holes and true Unicorns in a binary population \citep{Heger2003}, reducing the MS true Unicorn probability by as much as 10-25\% on average (for companions more massive than the MS star). This decrease becomes negligible when the ``dark'' companion is $\ge$3 times less luminous than the bright star}.
  \item {We observe that a young star cluster with a top-heavy IMF such as the Arches Cluster (modelled in this study) represents a more productive environment in searching for non-interacting compact objects.  We find a $9.6\pm3.8$\% MS true Unicorn probability at 4 Myrs for an Arches Cluster analogue modelled under a top-heavy IMF and 1.4 Z$_{\odot}$ metallicity (provided the companion is more massive than the MS star), which increases to $100.0\pm6.5$\% if the companion is confirmed to be at least 1.5 times less luminous than the bright star.}
  \item {In line with physical expectation, we find that a true Unicorn will tend to have a wider orbital separation than a false positive with the same period. We also find that for binaries with a non-interacting compact object companion more massive than the bright star, the orbital period will be <100 days $\sim$87\% of the time, confirming that a significant fraction of these objects should be found in short periods. Binaries with long orbital periods are more likely to be false positives.}
  \item {A single binary pair can change definition between a true Unicorn and a false positive at different points in its lifetime. We provide an example of the evolutionary path of such a binary in Section \ref{subsec:Evolution}.}
  \item{A significant fraction of true Unicorns, although non-interacting for most of their lives, will interact at some point and become detectable through x-rays (50-60\% for a Kroupa IMF, 55-80\% for a Hosek IMF). This stresses the importance of using a combination of radial velocity measurement, astrometry, and spectroscopy to locate these objects, given that roughly half of binaries with a bright star and a more massive BH/NS companion in relatively close orbits will never be detectable through x-ray emission during a 13 Gyr lifetime.}
\end{itemize}

In the search for non-interacting binaries containing a black hole or neutron star, radial velocity measurements and astrometry combined with spectroscopy and SED analysis constitutes a promising detection method. Only a small number of such objects have been discovered to date, and false positives have complicated the process. However, our models show that while binaries with non-interacting compact objects are indeed rare, there are conditions under which the probability of false positives decreases significantly. If the age of the bright star can be calculated through other means, our models show that binaries of certain types, ages, and orbital parameters are much more likely to contain a true a non-interacting compact object than a star acting as an imposter.

\section{Acknowledgements}

We thank our anonymous reviewers for carefully reviewing the manuscript and providing valuable comments. A.M.M. thanks Mark Reynolds and Dominick Rowan for helpful advice and suggestions. A.P.S. thanks Katelyn Breivik for helpful discussions regarding the {\tt COSMIC} code. A.M.M. acknowledges partial support from the Ohio State University Summer Undergraduate Research Program (SURP) in Astrophysics. A.P.S. acknowledges partial support from the President's Postdoctoral Scholarship by the Ohio State University and the Ohio Eminent Scholar Endowment. This research made use of {\tt COSMIC} \citep{Breivik2020}, {\tt Matplotlib} \citep{Hunter2007}, {\tt Numpy} \citep{vanderWalt2011}, {\tt Pandas} \citep{McKinney2010}, and {\tt Scipy} \citep{Virtanen2020}.

\section{Data Availability}

The output from all of the {\tt COSMIC} simulations can be provided upon request to the corresponding author.

\bibliographystyle{mnras}
\bibliography{references}

\begin{thebibliography}{}
\makeatletter
\relax
\def\mn@urlcharsother{\let\do\@makeother \do\$\do\&\do\#\do\^\do\_\do\%\do\~}
\def\mn@doi{\begingroup\mn@urlcharsother \@ifnextchar [ {\mn@doi@} {\mn@doi@[]}}
\def\mn@doi@[#1]#2{\def\@tempa{#1}\ifx\@tempa\@empty \href {http://dx.doi.org/#2} {doi:#2}\else \href {http://dx.doi.org/#2} {#1}\fi \endgroup}
\def\mn@eprint#1#2{\mn@eprint@#1:#2::\@nil}
\def\mn@eprint@arXiv#1{\href {http://arxiv.org/abs/#1} {{\tt arXiv:#1}}}
\def\mn@eprint@dblp#1{\href {http://dblp.uni-trier.de/rec/bibtex/#1.xml} {dblp:#1}}
\def\mn@eprint@#1:#2:#3:#4\@nil{\def\@tempa {#1}\def\@tempb {#2}\def\@tempc {#3}\ifx \@tempc \@empty \let \@tempc \@tempb \let \@tempb \@tempa \fi \ifx \@tempb \@empty \def\@tempb {arXiv}\fi \@ifundefined {mn@eprint@\@tempb}{\@tempb:\@tempc}{\expandafter \expandafter \csname mn@eprint@\@tempb\endcsname \expandafter{\@tempc}}}

\bibitem[\protect\citeauthoryear{{Abbott} et~al.,}{{Abbott} et~al.}{2016}]{Abbott2016}
{Abbott} B.~P.,  et~al., 2016, \mn@doi [\prl] {10.1103/PhysRevLett.116.061102}, \href {https://ui.adsabs.harvard.edu/abs/2016PhRvL.116f1102A} {116, 061102}

\bibitem[\protect\citeauthoryear{{Andrews}, {Breivik}  \& {Chatterjee}}{{Andrews} et~al.}{2019}]{Andrews2019}
{Andrews} J.~J.,  {Breivik} K.,   {Chatterjee} S.,  2019, \mn@doi [\apj] {10.3847/1538-4357/ab441f}, \href {https://ui.adsabs.harvard.edu/abs/2019ApJ...886...68A} {886, 68}

\bibitem[\protect\citeauthoryear{{Asplund}, {Grevesse}, {Sauval}  \& {Scott}}{{Asplund} et~al.}{2009}]{Asplund2009}
{Asplund} M.,  {Grevesse} N.,  {Sauval} A.~J.,   {Scott} P.,  2009, \mn@doi [\araa] {10.1146/annurev.astro.46.060407.145222}, \href {https://ui.adsabs.harvard.edu/abs/2009ARA&A..47..481A} {47, 481}

\bibitem[\protect\citeauthoryear{{Babel} \& {Montmerle}}{{Babel} \& {Montmerle}}{1997}]{Babel1997}
{Babel} J.,  {Montmerle} T.,  1997, \mn@doi [\apjl] {10.1086/310806}, \href {https://ui.adsabs.harvard.edu/abs/1997ApJ...485L..29B} {485, L29}

\bibitem[\protect\citeauthoryear{{Barnes}}{{Barnes}}{2003}]{Barnes2003}
{Barnes} S.~A.,  2003, \mn@doi [\apj] {10.1086/367639}, \href {https://ui.adsabs.harvard.edu/abs/2003ApJ...586..464B} {586, 464}

\bibitem[\protect\citeauthoryear{{Belczynski}, {Kalogera}, {Rasio}, {Taam}, {Zezas}, {Bulik}, {Maccarone}  \& {Ivanova}}{{Belczynski} et~al.}{2008}]{Belczynski2008}
{Belczynski} K.,  {Kalogera} V.,  {Rasio} F.~A.,  {Taam} R.~E.,  {Zezas} A.,  {Bulik} T.,  {Maccarone} T.~J.,   {Ivanova} N.,  2008, \mn@doi [\apjs] {10.1086/521026}, \href {https://ui.adsabs.harvard.edu/abs/2008ApJS..174..223B} {174, 223}

\bibitem[\protect\citeauthoryear{{Breivik}, {Chatterjee}  \& {Andrews}}{{Breivik} et~al.}{2019}]{Breivik2019CO}
{Breivik} K.,  {Chatterjee} S.,   {Andrews} J.~J.,  2019, \mn@doi [\apjl] {10.3847/2041-8213/ab21d3}, \href {https://ui.adsabs.harvard.edu/abs/2019ApJ...878L...4B} {878, L4}

\bibitem[\protect\citeauthoryear{{Breivik} et~al.,}{{Breivik} et~al.}{2020}]{Breivik2020}
{Breivik} K.,  et~al., 2020, \mn@doi [\apj] {10.3847/1538-4357/ab9d85}, \href {https://ui.adsabs.harvard.edu/abs/2020ApJ...898...71B} {898, 71}

\bibitem[\protect\citeauthoryear{{Caiazzo} et~al.,}{{Caiazzo} et~al.}{2021}]{Caiazzo2021}
{Caiazzo} I.,  et~al., 2021, \mn@doi [\nat] {10.1038/s41586-021-03615-y}, \href {https://ui.adsabs.harvard.edu/abs/2021Natur.595...39C} {595, 39}

\bibitem[\protect\citeauthoryear{{Charles} \& {Coe}}{{Charles} \& {Coe}}{2006}]{Charles&Coe2006}
{Charles} P.~A.,  {Coe} M.~J.,  2006, in , Vol.~39, Compact stellar X-ray sources.
pp 215--265

\bibitem[\protect\citeauthoryear{{Chawla}, {Chatterjee}, {Breivik}, {Moorthy}, {Andrews}  \& {Sanderson}}{{Chawla} et~al.}{2022}]{Chawla2022}
{Chawla} C.,  {Chatterjee} S.,  {Breivik} K.,  {Moorthy} C.~K.,  {Andrews} J.~J.,   {Sanderson} R.~E.,  2022, \mn@doi [\apj] {10.3847/1538-4357/ac60a5}, \href {https://ui.adsabs.harvard.edu/abs/2022ApJ...931..107C} {931, 107}

\bibitem[\protect\citeauthoryear{{Corral-Santana}, {Casares}, {Munoz-Darias}, {Bauer}, {Martinez-Pais}  \& {Russell}}{{Corral-Santana} et~al.}{2015}]{Corral-Santana2015}
{Corral-Santana} J.~M.,  {Casares} J.,  {Munoz-Darias} T.,  {Bauer} F.~E.,  {Martinez-Pais} I.~G.,   {Russell} D.~M.,  2015, VizieR Online Data Catalog, \href {https://ui.adsabs.harvard.edu/abs/2015yCat..35870061C} {pp J/A+A/587/A61}

\bibitem[\protect\citeauthoryear{{Dunn}, {Fender}, {K{\"o}rding}, {Belloni}  \& {Cabanac}}{{Dunn} et~al.}{2010}]{Dunn2010}
{Dunn} R.~J.~H.,  {Fender} R.~P.,  {K{\"o}rding} E.~G.,  {Belloni} T.,   {Cabanac} C.,  2010, \mn@doi [\mnras] {10.1111/j.1365-2966.2010.16114.x}, \href {https://ui.adsabs.harvard.edu/abs/2010MNRAS.403...61D} {403, 61}

\bibitem[\protect\citeauthoryear{{Duquennoy}, {Mayor}  \& {Halbwachs}}{{Duquennoy} et~al.}{1991}]{Duquennoy1991}
{Duquennoy} A.,  {Mayor} M.,   {Halbwachs} J.~L.,  1991, \aaps, \href {https://ui.adsabs.harvard.edu/abs/1991A&AS...88..281D} {88, 281}

\bibitem[\protect\citeauthoryear{{El-Badry}}{{El-Badry}}{2022}]{ElBadry2022BinaryZoo}
{El-Badry} K.,  2022, \mn@doi [\mnras] {10.1093/mnras/stac815}, \href {https://ui.adsabs.harvard.edu/abs/2022MNRAS.512.5620E} {512, 5620}

\bibitem[\protect\citeauthoryear{{El-Badry} \& {Burdge}}{{El-Badry} \& {Burdge}}{2022}]{El-Badry2022AnotherImposter}
{El-Badry} K.,  {Burdge} K.~B.,  2022, \mn@doi [\mnras] {10.1093/mnrasl/slab135}, \href {https://ui.adsabs.harvard.edu/abs/2022MNRAS.511L..24E} {511, 24}

\bibitem[\protect\citeauthoryear{{El-Badry} et~al.,}{{El-Badry} et~al.}{2018}]{El-Badry2018}
{El-Badry} K.,  et~al., 2018, \mn@doi [\mnras] {10.1093/mnras/sty240}, \href {https://ui.adsabs.harvard.edu/abs/2018MNRAS.476..528E} {476, 528}

\bibitem[\protect\citeauthoryear{{El-Badry}, {Burdge}  \& {Mr{\'o}z}}{{El-Badry} et~al.}{2022}]{El-Badry2022BHImposter}
{El-Badry} K.,  {Burdge} K.~B.,   {Mr{\'o}z} P.,  2022, \mn@doi [\mnras] {10.1093/mnras/stac274}, \href {https://ui.adsabs.harvard.edu/abs/2022MNRAS.511.3089E} {511, 3089}

\bibitem[\protect\citeauthoryear{{El-Badry} et~al.,}{{El-Badry} et~al.}{2023a}]{El-Badry2023}
{El-Badry} K.,  et~al., 2023a, \mn@doi [\mnras] {10.1093/mnras/stac3140}, \href {https://ui.adsabs.harvard.edu/abs/2023MNRAS.518.1057E} {518, 1057}

\bibitem[\protect\citeauthoryear{{El-Badry} et~al.,}{{El-Badry} et~al.}{2023b}]{El-Badry2023RG}
{El-Badry} K.,  et~al., 2023b, \mn@doi [\mnras] {10.1093/mnras/stad799}, \href {https://ui.adsabs.harvard.edu/abs/2023MNRAS.521.4323E} {521, 4323}

\bibitem[\protect\citeauthoryear{{El-Badry} et~al.,}{{El-Badry} et~al.}{2024a}]{ElBadry2024NScand}
{El-Badry} K.,  et~al., 2024a, \mn@doi [arXiv e-prints] {10.48550/arXiv.2405.00089}, \href {https://ui.adsabs.harvard.edu/abs/2024arXiv240500089E} {p. arXiv:2405.00089}

\bibitem[\protect\citeauthoryear{{El-Badry} et~al.,}{{El-Badry} et~al.}{2024b}]{ElBadry2024NS1}
{El-Badry} K.,  et~al., 2024b, \mn@doi [The Open Journal of Astrophysics] {10.33232/001c.116675}, \href {https://ui.adsabs.harvard.edu/abs/2024OJAp....7E..27E} {7, 27}

\bibitem[\protect\citeauthoryear{{Feldmeier}, {Puls}  \& {Pauldrach}}{{Feldmeier} et~al.}{1997}]{Feldmeier1997}
{Feldmeier} A.,  {Puls} J.,   {Pauldrach} A.~W.~A.,  1997, \aap, \href {https://ui.adsabs.harvard.edu/abs/1997A&A...322..878F} {322, 878}

\bibitem[\protect\citeauthoryear{{Finn}}{{Finn}}{1994}]{Finn1994}
{Finn} L.~S.,  1994, \mn@doi [\prl] {10.1103/PhysRevLett.73.1878}, \href {https://ui.adsabs.harvard.edu/abs/1994PhRvL..73.1878F} {73, 1878}

\bibitem[\protect\citeauthoryear{{Gaia Collaboration} et~al.,}{{Gaia Collaboration} et~al.}{2024}]{GaiaCollab2024}
{Gaia Collaboration} et~al., 2024, \mn@doi [arXiv e-prints] {10.48550/arXiv.2404.10486}, \href {https://ui.adsabs.harvard.edu/abs/2024arXiv240410486G} {p. arXiv:2404.10486}

\bibitem[\protect\citeauthoryear{{Gallego-Calvente} et~al.,}{{Gallego-Calvente} et~al.}{2021}]{Gallego2021}
{Gallego-Calvente} A.~T.,  et~al., 2021, \mn@doi [\aap] {10.1051/0004-6361/202039172}, \href {https://ui.adsabs.harvard.edu/abs/2021A&A...647A.110G} {647, A110}

\bibitem[\protect\citeauthoryear{{Giacconi}, {Gursky}  \& {Waters}}{{Giacconi} et~al.}{1964}]{Giacconi1964}
{Giacconi} R.,  {Gursky} H.,   {Waters} J.~R.,  1964, \mn@doi [\nat] {10.1038/204981a0}, \href {https://ui.adsabs.harvard.edu/abs/1964Natur.204..981G} {204, 981}

\bibitem[\protect\citeauthoryear{{Gilfanov}, {Fabbiano}, {Lehmer}  \& {Zezas}}{{Gilfanov} et~al.}{2022}]{Gilfanov2023}
{Gilfanov} M.,  {Fabbiano} G.,  {Lehmer} B.,   {Zezas} A.,  2022, in , Handbook of X-ray and Gamma-ray Astrophysics.
p.~105, \mn@doi{10.1007/978-981-16-4544-0_108-1}

\bibitem[\protect\citeauthoryear{{Guseinov} \& {Zel'dovich}}{{Guseinov} \& {Zel'dovich}}{1966}]{Guseinov1966}
{Guseinov} O.~K.,  {Zel'dovich} Y.~B.,  1966, \sovast, \href {https://ui.adsabs.harvard.edu/abs/1966SvA....10..251G} {10, 251}

\bibitem[\protect\citeauthoryear{{Haensel}, {Zdunik}  \& {Douchin}}{{Haensel} et~al.}{2002}]{Haensel2002}
{Haensel} P.,  {Zdunik} J.~L.,   {Douchin} F.,  2002, \mn@doi [\aap] {10.1051/0004-6361:20020131}, \href {https://ui.adsabs.harvard.edu/abs/2002A&A...385..301H} {385, 301}

\bibitem[\protect\citeauthoryear{{Heger}, {Fryer}, {Woosley}, {Langer}  \& {Hartmann}}{{Heger} et~al.}{2003}]{Heger2003}
{Heger} A.,  {Fryer} C.~L.,  {Woosley} S.~E.,  {Langer} N.,   {Hartmann} D.~H.,  2003, \mn@doi [\apj] {10.1086/375341}, \href {https://ui.adsabs.harvard.edu/abs/2003ApJ...591..288H} {591, 288}

\bibitem[\protect\citeauthoryear{{Hirai} \& {Mandel}}{{Hirai} \& {Mandel}}{2021}]{Hirai2021}
{Hirai} R.,  {Mandel} I.,  2021, \mn@doi [\pasa] {10.1017/pasa.2021.53}, \href {https://ui.adsabs.harvard.edu/abs/2021PASA...38...56H} {38, e056}

\bibitem[\protect\citeauthoryear{{Hjellming} \& {Webbink}}{{Hjellming} \& {Webbink}}{1987}]{Hjellming1987}
{Hjellming} M.~S.,  {Webbink} R.~F.,  1987, \mn@doi [\apj] {10.1086/165412}, \href {https://ui.adsabs.harvard.edu/abs/1987ApJ...318..794H} {318, 794}

\bibitem[\protect\citeauthoryear{{Hoang}, {Naoz}, {Kocsis}, {Rasio}  \& {Dosopoulou}}{{Hoang} et~al.}{2018}]{Hoang2018}
{Hoang} B.-M.,  {Naoz} S.,  {Kocsis} B.,  {Rasio} F.~A.,   {Dosopoulou} F.,  2018, \mn@doi [\apj] {10.3847/1538-4357/aaafce}, \href {https://ui.adsabs.harvard.edu/abs/2018ApJ...856..140H} {856, 140}

\bibitem[\protect\citeauthoryear{{Hosek}}{{Hosek}}{2019}]{Hosek2019}
{Hosek} Matthew~W. J.,  2019, \mn@doi [\apj] {10.3847/1538-4357/aaef90}, \href {https://ui.adsabs.harvard.edu/abs/2019ApJ...870...44H} {870, 44}

\bibitem[\protect\citeauthoryear{{Hunt}, {Gallo}, {Chandar}, {Johns Mulia}, {Mok}, {Prestwich}  \& {Liu}}{{Hunt} et~al.}{2021}]{Hunt2021}
{Hunt} Q.,  {Gallo} E.,  {Chandar} R.,  {Johns Mulia} P.,  {Mok} A.,  {Prestwich} A.,   {Liu} S.,  2021, \mn@doi [\apj] {10.3847/1538-4357/abe531}, \href {https://ui.adsabs.harvard.edu/abs/2021ApJ...912...31H} {912, 31}

\bibitem[\protect\citeauthoryear{{Hunter}}{{Hunter}}{2007}]{Hunter2007}
{Hunter} J.~D.,  2007, \mn@doi [Computing in Science and Engineering] {10.1109/MCSE.2007.55}, \href {https://ui.adsabs.harvard.edu/abs/2007CSE.....9...90H} {9, 90}

\bibitem[\protect\citeauthoryear{{Hurley}, {Pols}  \& {Tout}}{{Hurley} et~al.}{2000}]{Hurley2000}
{Hurley} J.~R.,  {Pols} O.~R.,   {Tout} C.~A.,  2000, \mn@doi [\mnras] {10.1046/j.1365-8711.2000.03426.x}, \href {https://ui.adsabs.harvard.edu/abs/2000MNRAS.315..543H} {315, 543}

\bibitem[\protect\citeauthoryear{{Hurley}, {Tout}  \& {Pols}}{{Hurley} et~al.}{2002}]{Hurley2002}
{Hurley} J.~R.,  {Tout} C.~A.,   {Pols} O.~R.,  2002, \mn@doi [\mnras] {10.1046/j.1365-8711.2002.05038.x}, \href {https://ui.adsabs.harvard.edu/abs/2002MNRAS.329..897H} {329, 897}

\bibitem[\protect\citeauthoryear{{Irrgang}, {Geier}, {Kreuzer}, {Pelisoli}  \& {Heber}}{{Irrgang} et~al.}{2020}]{Irrgang2020}
{Irrgang} A.,  {Geier} S.,  {Kreuzer} S.,  {Pelisoli} I.,   {Heber} U.,  2020, \mn@doi [\aap] {10.1051/0004-6361/201937343}, \href {https://ui.adsabs.harvard.edu/abs/2020A&A...633L...5I} {633, L5}

\bibitem[\protect\citeauthoryear{{Ivanova} \& {Taam}}{{Ivanova} \& {Taam}}{2003}]{Ivanova2003}
{Ivanova} N.,  {Taam} R.~E.,  2003, \mn@doi [\apj] {10.1086/379192}, \href {https://ui.adsabs.harvard.edu/abs/2003ApJ...599..516I} {599, 516}

\bibitem[\protect\citeauthoryear{{Jayasinghe} et~al.,}{{Jayasinghe} et~al.}{2021}]{Jayasinghe2021Unicorn}
{Jayasinghe} T.,  et~al., 2021, \mn@doi [\mnras] {10.1093/mnras/stab907}, \href {https://ui.adsabs.harvard.edu/abs/2021MNRAS.504.2577J} {504, 2577}

\bibitem[\protect\citeauthoryear{{Jayasinghe} et~al.,}{{Jayasinghe} et~al.}{2022}]{Jayasinghe2022Giraffe}
{Jayasinghe} T.,  et~al., 2022, \mn@doi [\mnras] {10.1093/mnras/stac2187}, \href {https://ui.adsabs.harvard.edu/abs/2022MNRAS.516.5945J} {516, 5945}

\bibitem[\protect\citeauthoryear{{Jordan}}{{Jordan}}{2008}]{Jordan2008}
{Jordan} S.,  2008, \mn@doi [Astronomische Nachrichten] {10.1002/asna.200811065}, \href {https://ui.adsabs.harvard.edu/abs/2008AN....329..875J} {329, 875}

\bibitem[\protect\citeauthoryear{{Kalogera} \& {Baym}}{{Kalogera} \& {Baym}}{1996}]{Kalogera1996}
{Kalogera} V.,  {Baym} G.,  1996, \mn@doi [\apjl] {10.1086/310296}, \href {https://ui.adsabs.harvard.edu/abs/1996ApJ...470L..61K} {470, L61}

\bibitem[\protect\citeauthoryear{{Kaplan} et~al.,}{{Kaplan} et~al.}{2014}]{Kaplan2014}
{Kaplan} D.~L.,  et~al., 2014, \mn@doi [\apj] {10.1088/0004-637X/780/2/167}, \href {https://ui.adsabs.harvard.edu/abs/2014ApJ...780..167K} {780, 167}

\bibitem[\protect\citeauthoryear{{Kawka} \& {Vennes}}{{Kawka} \& {Vennes}}{2009}]{Kawka2009}
{Kawka} A.,  {Vennes} S.,  2009, \mn@doi [\aap] {10.1051/0004-6361/200912954}, \href {https://ui.adsabs.harvard.edu/abs/2009A&A...506L..25K} {506, L25}

\bibitem[\protect\citeauthoryear{{Kim}, {Figer}, {Kudritzki}  \& {Najarro}}{{Kim} et~al.}{2006}]{Kim2006}
{Kim} S.~S.,  {Figer} D.~F.,  {Kudritzki} R.~P.,   {Najarro} F.,  2006, \mn@doi [\apjl] {10.1086/510529}, \href {https://ui.adsabs.harvard.edu/abs/2006ApJ...653L.113K} {653, L113}

\bibitem[\protect\citeauthoryear{{King}}{{King}}{1995}]{King1995}
{King} A.,  1995, in X-ray Binaries. pp 419--456

\bibitem[\protect\citeauthoryear{{Kozai}}{{Kozai}}{1962}]{Kozai1962}
{Kozai} Y.,  1962, \mn@doi [\aj] {10.1086/108790}, \href {https://ui.adsabs.harvard.edu/abs/1962AJ.....67..591K} {67, 591}

\bibitem[\protect\citeauthoryear{{Krause} et~al.,}{{Krause} et~al.}{2020}]{Krause2020}
{Krause} M. G.~H.,  et~al., 2020, \mn@doi [\ssr] {10.1007/s11214-020-00689-4}, \href {https://ui.adsabs.harvard.edu/abs/2020SSRv..216...64K} {216, 64}

\bibitem[\protect\citeauthoryear{{Kroupa}, {Tout}  \& {Gilmore}}{{Kroupa} et~al.}{1993}]{Kroupa1993}
{Kroupa} P.,  {Tout} C.~A.,   {Gilmore} G.,  1993, \mn@doi [\mnras] {10.1093/mnras/262.3.545}, \href {https://ui.adsabs.harvard.edu/abs/1993MNRAS.262..545K} {262, 545}

\bibitem[\protect\citeauthoryear{{LIGO} et~al.,}{{LIGO} et~al.}{2021}]{ligo2021}
{LIGO} et~al., 2021, \mn@doi [arXiv e-prints] {10.48550/arXiv.2111.03606}, \href {https://ui.adsabs.harvard.edu/abs/2021arXiv211103606T} {p. arXiv:2111.03606}

\bibitem[\protect\citeauthoryear{{Lattimer} \& {Prakash}}{{Lattimer} \& {Prakash}}{2004}]{Lattimer2004}
{Lattimer} J.~M.,  {Prakash} M.,  2004, \mn@doi [Science] {10.1126/science.1090720}, \href {https://ui.adsabs.harvard.edu/abs/2004Sci...304..536L} {304, 536}

\bibitem[\protect\citeauthoryear{{Lidov}}{{Lidov}}{1962}]{Lidov1962}
{Lidov} M.~L.,  1962, \mn@doi [\planss] {10.1016/0032-0633(62)90129-0}, \href {https://ui.adsabs.harvard.edu/abs/1962P&SS....9..719L} {9, 719}

\bibitem[\protect\citeauthoryear{{Liotine}, {Zevin}, {Berry}, {Doctor}  \& {Kalogera}}{{Liotine} et~al.}{2023}]{Liotine2023}
{Liotine} C.,  {Zevin} M.,  {Berry} C. P.~L.,  {Doctor} Z.,   {Kalogera} V.,  2023, \mn@doi [\apj] {10.3847/1538-4357/acb8b2}, \href {https://ui.adsabs.harvard.edu/abs/2023ApJ...946....4L} {946, 4}

\bibitem[\protect\citeauthoryear{{Liu} et~al.,}{{Liu} et~al.}{2019}]{Liu2019}
{Liu} J.,  et~al., 2019, \mn@doi [\nat] {10.1038/s41586-019-1766-2}, \href {https://ui.adsabs.harvard.edu/abs/2019Natur.575..618L} {575, 618}

\bibitem[\protect\citeauthoryear{{Lorimer} et~al.,}{{Lorimer} et~al.}{2021}]{Lorimier2021}
{Lorimer} D.~R.,  et~al., 2021, \mn@doi [\mnras] {10.1093/mnras/stab2474}, \href {https://ui.adsabs.harvard.edu/abs/2021MNRAS.507.5303L} {507, 5303}

\bibitem[\protect\citeauthoryear{{Martins}, {Hillier}, {Paumard}, {Eisenhauer}, {Ott}  \& {Genzel}}{{Martins} et~al.}{2008}]{Martins2008}
{Martins} F.,  {Hillier} D.~J.,  {Paumard} T.,  {Eisenhauer} F.,  {Ott} T.,   {Genzel} R.,  2008, \mn@doi [\aap] {10.1051/0004-6361:20078469}, \href {https://ui.adsabs.harvard.edu/abs/2008A&A...478..219M} {478, 219}

\bibitem[\protect\citeauthoryear{{Mashian} \& {Loeb}}{{Mashian} \& {Loeb}}{2017}]{Mashian2017}
{Mashian} N.,  {Loeb} A.,  2017, \mn@doi [\mnras] {10.1093/mnras/stx1410}, \href {https://ui.adsabs.harvard.edu/abs/2017MNRAS.470.2611M} {470, 2611}

\bibitem[\protect\citeauthoryear{{McKinney}}{{McKinney}}{2010}]{McKinney2010}
{McKinney} W.,  2010, in {S}t\'efan van~der {W}alt {J}arrod {M}illman eds, {P}roceedings of the 9th {P}ython in {S}cience {C}onference. pp 56 -- 61, \mn@doi{10.25080/Majora-92bf1922-00a}

\bibitem[\protect\citeauthoryear{{Mineo}, {Gilfanov}  \& {Sunyaev}}{{Mineo} et~al.}{2012}]{Mineo2012}
{Mineo} S.,  {Gilfanov} M.,   {Sunyaev} R.,  2012, \mn@doi [\mnras] {10.1111/j.1365-2966.2011.19862.x}, \href {https://ui.adsabs.harvard.edu/abs/2012MNRAS.419.2095M} {419, 2095}

\bibitem[\protect\citeauthoryear{{Misra} et~al.,}{{Misra} et~al.}{2023}]{Misra2023}
{Misra} D.,  et~al., 2023, \mn@doi [\aap] {10.1051/0004-6361/202244929}, \href {https://ui.adsabs.harvard.edu/abs/2023A&A...672A..99M} {672, A99}

\bibitem[\protect\citeauthoryear{{Naoz}}{{Naoz}}{2016}]{Naoz2016}
{Naoz} S.,  2016, \mn@doi [\araa] {10.1146/annurev-astro-081915-023315}, \href {https://ui.adsabs.harvard.edu/abs/2016ARA&A..54..441N} {54, 441}

\bibitem[\protect\citeauthoryear{{Naoz} \& {Fabrycky}}{{Naoz} \& {Fabrycky}}{2014}]{Naoz2014}
{Naoz} S.,  {Fabrycky} D.~C.,  2014, \mn@doi [\apj] {10.1088/0004-637X/793/2/137}, \href {https://ui.adsabs.harvard.edu/abs/2014ApJ...793..137N} {793, 137}

\bibitem[\protect\citeauthoryear{{Overbeck} \& {Tananbaum}}{{Overbeck} \& {Tananbaum}}{1968}]{Overbeck1968}
{Overbeck} J.~W.,  {Tananbaum} H.~D.,  1968, \mn@doi [\apj] {10.1086/149714}, \href {https://ui.adsabs.harvard.edu/abs/1968ApJ...153..899O} {153, 899}

\bibitem[\protect\citeauthoryear{{Pfahl}, {Rappaport}, {Podsiadlowski}  \& {Spruit}}{{Pfahl} et~al.}{2002}]{Pfahl2002}
{Pfahl} E.,  {Rappaport} S.,  {Podsiadlowski} P.,   {Spruit} H.,  2002, \mn@doi [\apj] {10.1086/340794}, \href {https://ui.adsabs.harvard.edu/abs/2002ApJ...574..364P} {574, 364}

\bibitem[\protect\citeauthoryear{{Pfahl}, {Rappaport}  \& {Podsiadlowski}}{{Pfahl} et~al.}{2003}]{Pfahl2003}
{Pfahl} E.,  {Rappaport} S.,   {Podsiadlowski} P.,  2003, \mn@doi [\apj] {10.1086/378632}, \href {https://ui.adsabs.harvard.edu/abs/2003ApJ...597.1036P} {597, 1036}

\bibitem[\protect\citeauthoryear{{Podsiadlowski}, {Rappaport}  \& {Pfahl}}{{Podsiadlowski} et~al.}{2002}]{Podsiadlowski2002}
{Podsiadlowski} P.,  {Rappaport} S.,   {Pfahl} E.~D.,  2002, \mn@doi [\apj] {10.1086/324686}, \href {https://ui.adsabs.harvard.edu/abs/2002ApJ...565.1107P} {565, 1107}

\bibitem[\protect\citeauthoryear{{Podsiadlowski}, {Rappaport}  \& {Han}}{{Podsiadlowski} et~al.}{2003}]{Podsiadlowski2003}
{Podsiadlowski} P.,  {Rappaport} S.,   {Han} Z.,  2003, \mn@doi [\mnras] {10.1046/j.1365-8711.2003.06464.x}, \href {https://ui.adsabs.harvard.edu/abs/2003MNRAS.341..385P} {341, 385}

\bibitem[\protect\citeauthoryear{{Raghavan} et~al.,}{{Raghavan} et~al.}{2010}]{Raghavan2010}
{Raghavan} D.,  et~al., 2010, \mn@doi [\apjs] {10.1088/0067-0049/190/1/1}, \href {https://ui.adsabs.harvard.edu/abs/2010ApJS..190....1R} {190, 1}

\bibitem[\protect\citeauthoryear{{Rauw} et~al.,}{{Rauw} et~al.}{2015}]{Rauw2015}
{Rauw} G.,  et~al., 2015, \mn@doi [\apjs] {10.1088/0067-0049/221/1/1}, \href {https://ui.adsabs.harvard.edu/abs/2015ApJS..221....1R} {221, 1}

\bibitem[\protect\citeauthoryear{{Romero-Shaw}, {Hirai}, {Bahramian}, {Willcox}  \& {Mandel}}{{Romero-Shaw} et~al.}{2023}]{Romero-Shaw2023}
{Romero-Shaw} I.,  {Hirai} R.,  {Bahramian} A.,  {Willcox} R.,   {Mandel} I.,  2023, \mn@doi [\mnras] {10.1093/mnras/stad1732}, \href {https://ui.adsabs.harvard.edu/abs/2023MNRAS.524..245R} {524, 245}

\bibitem[\protect\citeauthoryear{{Rose}, {Naoz}, {Sari}  \& {Linial}}{{Rose} et~al.}{2023}]{Rose2023}
{Rose} S.~C.,  {Naoz} S.,  {Sari} R.,   {Linial} I.,  2023, \mn@doi [\apj] {10.3847/1538-4357/acee75}, \href {https://ui.adsabs.harvard.edu/abs/2023ApJ...955...30R} {955, 30}

\bibitem[\protect\citeauthoryear{{Sabhahit}, {Vink}, {Higgins}  \& {Sander}}{{Sabhahit} et~al.}{2022}]{Sabhahit2022}
{Sabhahit} G.~N.,  {Vink} J.~S.,  {Higgins} E.~R.,   {Sander} A. A.~C.,  2022, \mn@doi [\mnras] {10.1093/mnras/stac1410}, \href {https://ui.adsabs.harvard.edu/abs/2022MNRAS.514.3736S} {514, 3736}

\bibitem[\protect\citeauthoryear{{Salpeter}}{{Salpeter}}{1955}]{Salpeter1955}
{Salpeter} E.~E.,  1955, \mn@doi [\apj] {10.1086/145971}, \href {https://ui.adsabs.harvard.edu/abs/1955ApJ...121..161S} {121, 161}

\bibitem[\protect\citeauthoryear{{Sguera}, {Sidoli}, {Bird}  \& {La Palombara}}{{Sguera} et~al.}{2023}]{Sguera2023}
{Sguera} V.,  {Sidoli} L.,  {Bird} A.~J.,   {La Palombara} N.,  2023, \mn@doi [\mnras] {10.1093/mnras/stad1494}, \href {https://ui.adsabs.harvard.edu/abs/2023MNRAS.523.1192S} {523, 1192}

\bibitem[\protect\citeauthoryear{{Shakura} \& {Sunyaev}}{{Shakura} \& {Sunyaev}}{1973}]{Shakura1973}
{Shakura} N.~I.,  {Sunyaev} R.~A.,  1973, \aap, \href {https://ui.adsabs.harvard.edu/abs/1973A&A....24..337S} {24, 337}

\bibitem[\protect\citeauthoryear{{Stevens}, {Blondin}  \& {Pollock}}{{Stevens} et~al.}{1992}]{Stevens1992}
{Stevens} I.~R.,  {Blondin} J.~M.,   {Pollock} A.~M.~T.,  1992, \mn@doi [\apj] {10.1086/171013}, \href {https://ui.adsabs.harvard.edu/abs/1992ApJ...386..265S} {386, 265}

\bibitem[\protect\citeauthoryear{{Strobel}, {Schaab}  \& {Weigel}}{{Strobel} et~al.}{1999}]{Strobel1999}
{Strobel} K.,  {Schaab} C.,   {Weigel} M.~K.,  1999, \mn@doi [\aap] {10.48550/arXiv.astro-ph/9908132}, \href {https://ui.adsabs.harvard.edu/abs/1999A&A...350..497S} {350, 497}

\bibitem[\protect\citeauthoryear{{Tanaka} \& {Shibazaki}}{{Tanaka} \& {Shibazaki}}{1996}]{Tanaka&Shibazaki1996}
{Tanaka} Y.,  {Shibazaki} N.,  1996, \mn@doi [\araa] {10.1146/annurev.astro.34.1.607}, \href {https://ui.adsabs.harvard.edu/abs/1996ARA&A..34..607T} {34, 607}

\bibitem[\protect\citeauthoryear{{Tananbaum}, {Gursky}, {Kellogg}, {Levinson}, {Schreier}  \& {Giacconi}}{{Tananbaum} et~al.}{1972}]{Tananbaum1972}
{Tananbaum} H.,  {Gursky} H.,  {Kellogg} E.~M.,  {Levinson} R.,  {Schreier} E.,   {Giacconi} R.,  1972, \mn@doi [\apjl] {10.1086/180968}, \href {https://ui.adsabs.harvard.edu/abs/1972ApJ...174L.143T} {174, L143}

\bibitem[\protect\citeauthoryear{{Tauris} \& {van den Heuvel}}{{Tauris} \& {van den Heuvel}}{2006}]{Tauris2006}
{Tauris} T.~M.,  {van den Heuvel} E.~P.~J.,  2006, in , Vol.~39, Compact stellar X-ray sources.
pp 623--665, \mn@doi{10.48550/arXiv.astro-ph/0303456}

\bibitem[\protect\citeauthoryear{{Thompson} et~al.,}{{Thompson} et~al.}{2019}]{Thompson2019}
{Thompson} T.~A.,  et~al., 2019, \mn@doi [Science] {10.1126/science.aau4005}, \href {https://ui.adsabs.harvard.edu/abs/2019Sci...366..637T} {366, 637}

\bibitem[\protect\citeauthoryear{{Trimble} \& {Thorne}}{{Trimble} \& {Thorne}}{1969}]{Trimble1969}
{Trimble} V.~L.,  {Thorne} K.~S.,  1969, \mn@doi [\apj] {10.1086/150032}, \href {https://ui.adsabs.harvard.edu/abs/1969ApJ...156.1013T} {156, 1013}

\bibitem[\protect\citeauthoryear{{Virtanen} et~al.,}{{Virtanen} et~al.}{2020}]{Virtanen2020}
{Virtanen} P.,  et~al., 2020, \mn@doi [Nature Methods] {10.1038/s41592-019-0686-2}, \href {https://ui.adsabs.harvard.edu/abs/2020NatMe..17..261V} {17, 261}

\bibitem[\protect\citeauthoryear{{Weller} \& {Johnson}}{{Weller} \& {Johnson}}{2023}]{Weller2023}
{Weller} M.~K.,  {Johnson} J.~A.,  2023, \mn@doi [\mnras] {10.1093/mnras/stad181}, \href {https://ui.adsabs.harvard.edu/abs/2023MNRAS.520..935W} {520, 935}

\bibitem[\protect\citeauthoryear{{Wiktorowicz}, {Lu}, {Wyrzykowski}, {Zhang}, {Liu}, {Justham}  \& {Belczynski}}{{Wiktorowicz} et~al.}{2020}]{Wiktorowicz2020}
{Wiktorowicz} G.,  {Lu} Y.,  {Wyrzykowski} {\L}.,  {Zhang} H.,  {Liu} J.,  {Justham} S.,   {Belczynski} K.,  2020, \mn@doi [\apj] {10.3847/1538-4357/abc699}, \href {https://ui.adsabs.harvard.edu/abs/2020ApJ...905..134W} {905, 134}

\bibitem[\protect\citeauthoryear{{Yamaguchi}, {Kawanaka}, {Bulik}  \& {Piran}}{{Yamaguchi} et~al.}{2018}]{Yamaguchi2018}
{Yamaguchi} M.~S.,  {Kawanaka} N.,  {Bulik} T.,   {Piran} T.,  2018, \mn@doi [\apj] {10.3847/1538-4357/aac5ec}, \href {https://ui.adsabs.harvard.edu/abs/2018ApJ...861...21Y} {861, 21}

\bibitem[\protect\citeauthoryear{{Yan} et~al.,}{{Yan} et~al.}{2021}]{Yan2021}
{Yan} Z.,  et~al., 2021, \mn@doi [\apj] {10.3847/1538-4357/ac25eb}, \href {https://ui.adsabs.harvard.edu/abs/2021ApJ...921..120Y} {921, 120}

\bibitem[\protect\citeauthoryear{{Yudin}, {Razinkova}  \& {Blinnikov}}{{Yudin} et~al.}{2020}]{Yudin2020}
{Yudin} A.~V.,  {Razinkova} T.~L.,   {Blinnikov} S.~I.,  2020, \mn@doi [Astronomy Letters] {10.1134/S1063773719120077}, \href {https://ui.adsabs.harvard.edu/abs/2020AstL...45..847Y} {45, 847}

\bibitem[\protect\citeauthoryear{{de Kool}}{{de Kool}}{1990}]{DeKool1990}
{de Kool} M.,  1990, \mn@doi [\apj] {10.1086/168974}, \href {https://ui.adsabs.harvard.edu/abs/1990ApJ...358..189D} {358, 189}

\bibitem[\protect\citeauthoryear{{ud-Doula} \& {Owocki}}{{ud-Doula} \& {Owocki}}{2002}]{ud-Doula2002}
{ud-Doula} A.,  {Owocki} S.~P.,  2002, \mn@doi [\apj] {10.1086/341543}, \href {https://ui.adsabs.harvard.edu/abs/2002ApJ...576..413U} {576, 413}

\bibitem[\protect\citeauthoryear{{van der Walt}, {Colbert}  \& {Varoquaux}}{{van der Walt} et~al.}{2011}]{vanderWalt2011}
{van der Walt} S.,  {Colbert} S.~C.,   {Varoquaux} G.,  2011, \mn@doi [Computing in Science and Engineering] {10.1109/MCSE.2011.37}, \href {https://ui.adsabs.harvard.edu/abs/2011CSE....13b..22V} {13, 22}

\makeatother
\end{thebibliography}

\appendix

\section{Additional COSMIC Input Parameters}\label{sec:appendixA}

\begin{table*}
    \centering
    \caption{General input physics parameters for {\tt COSMIC} version $3.4.0$, as detailed in \citet{Breivik2020}. For our study we have used {\tt COSMIC}'s default values, which appear in this table.}
    \label{tab:tableA1}
    \begin{tabular}{c|c|l}
    Flag & Value & Description\\
    \hline
    acc\_lim & -1 & Limits mass accreted during RLOF: 10x’s thermal rate of the accretor for MS/HG/CHeB; unlimited for GB/EAGB/AGB\\
    acc2 & 1.5 & Bondi-Hoyle wind accretion factor (mean wind accretion rate onto the secondary is proportional to this value)\\
    aic & 1 & Reduces kick strengths for accretion induced collapse SN according to \textit{sigmadiv}\\
    alpha1 & 1 & Common-envelope efficiency parameter, scales efficiency of transferring orbital energy to the envelope\\
    bconst & 3000 & Sets the magnetic field decay timescale for pulsars\\
    bdecayfac & 1 & Activates inverse decay model for accretion induced field decay\\
    beta & -1 & Wind velocity factor, scales with $v_{wind}^{2}$\\
    bhflag & 1 & Sets a fallback-modulated kick model for applying SN kicks to BHs\\
    bhms\_coll\_flag & 0 & Requires that in a BH/star collision, the star is  destroyed\\
    bhsigmafrac & 1 & Sets a fractional modification which scales down \textit{sigma} for BHs\\
    bhspinflag & 0 & Defines BH spin after formation, sets all values to match \textit{bhspinmag}\\
    bhspinmag & 0 & Sets the spin of all BHs, or the upper limit of BH spin distribution\\
    bwind & 0 & Binary enhanced mass loss parameter\\
    ceflag & 0 & Selects \citet{DeKool1990} model to set initial orbital energy, using the total mass of the stars instead of core mass\\
    cehestarflag & 0 & Removes fitting formulae for evolving RLOF systems with a He star donor and compact object accretor\\
    cekickflag & 2 & Selects post-CE values for mass and separation in the event that a supernova kick needs to be applied during CE\\
    cemergeflag & 0 & Stars do not automatically merge when beginning a CE without a distinct core-envelope boundary\\
    ck & 1000 & Sets the magnetic field decay timescale for pulsars\\
    don\_lim & -1 & Determines the rate of donor star mass loss through RLOF\\
    ecsn & 2.25 & Allows for electron capture SNe, sets the maximum He-star mass (at core helium depletion) that will result in an ECSN\\
    ecsn\_mlow & 1.6 & Sets the low end of the ECSN mass range\\
    eddfac & 1 & Eddington limit factor for mass transfer\\
    eddlimflag & 0 & Signals that mass loss does not depend on metallicity for stars near the Eddington limit\\
    epsnov & 0.001 & Fraction of accreted matter retained in a nova eruption\\
    gamma & -2 & During RLOF at super-Eddington mass transfer rates, material is lost if it is a wind from the secondary star\\
    grflag & 1 & Turns on orbital decay due to gravitational wave emission\\
    hewind & 0.5 & Helium star mass loss parameter (10$^{-13}$ \textit{hewind} L$^{2/3}$)\\
    htpmb & 1 & Activates \citet{Ivanova2003} model for magnetic braking\\
    ifflag & 0 & Activates initial-final white dwarf mass relation\\
    kickflag & 0 & Natal kicks drawn from a bimodal distribution; FeCCSN kick determined by \textit{sigma}, ECSN/USSN kick by \textit{sigmadiv}\\
    lambdaf & 0 & Requires no extra ionization energy goes into ejecting envelope during CE evolution\\
    mxns & 3 & Sets the boundary between the maximum NS mass and the minimum BH mass\\
    neta & 0.5 & Reimers mass-loss coefficient\\
    pisn & 45 & Allows for pulsational pair instability SNe, sets the maximum mass of the allowed remnant\\
    polar\_kick\_angle & 90 & Sets the opening angle of the SN kick relative to the pole of the exploding star, where \textit{90} gives fully isotropic kicks\\
    qcflag & 1 & Sets \citet{Hjellming1987} model for RGB, critical mass ratios for onset of unstable RLOF mass transfer/CE \\
    rejuv\_fac & 1 & Sets the mixing factor in MS star collisions\\
    rejuvflag & 0 & Selects no modifications to the original prescription for mixing of MS stars\\
    rembar\_massloss & 0.5 & Determines prescriptions for mass conversion due to neutrino emission during the collapse of the proto-compact object\\
    remnantflag & 4 & Selects a delayed remnant mass prescription, filling the mass gap between NS and BH\\
    sigma & 265 & Sets dispersion in the Maxwellian for the SN kick velocity in km/s\\
    sigmadiv & -20 & Sets ECSN kicks to be drawn from a Maxwellian distribution with dispersion given by this value\\
    ST\_cr & 1 & Follows \citet{Belczynski2008} model for convective vs radiative boundaries\\
    ST\_tide & 1 & Activates \citet{Belczynski2008} setup for tides\\
    tflag & 1 & Activates tidal circularization\\
    ussn & 0 & Reduces kicks according to \textit{sigmadiv} selection for ultra-stripped SNe\\
    wdflag & 1 & Activates a modified cooling law\\
    windflag & 3 & Wind mass loss is metallicity dependent for O/B, W-R stars, LBV-like for giants beyond Humphreys-Davidson limit\\
    xi & 1 & Fraction of angular momentum lost via winds from primary that transfers to the spin angular momentum of secondary\\
    
    \end{tabular}

\end{table*}

{\tt COSMIC} is a binary population synthesis and evolution code based on the older {\tt BSE} code \citep{Hurley2002}. It greatly expands upon the functionalities of {\tt BSE} by including more detailed stellar and binary interaction processes, and allowing for the evolution of more massive stars.

For this work we used {\tt COSMIC} version $3.4.0$ to evolve the synthetic binary pairs within our stellar populations. The general input physics parameters for {\tt COSMIC} are outlined in \citet{Breivik2020}. For our study we have used {\tt COSMIC}'s default values, which we have listed in Table \ref{tab:tableA1}.

\section{Evolutionary Results for Hosek IMF Population}\label{sec:appendixB}

\begin{figure*}
     \centering
     \includegraphics[width=\linewidth]{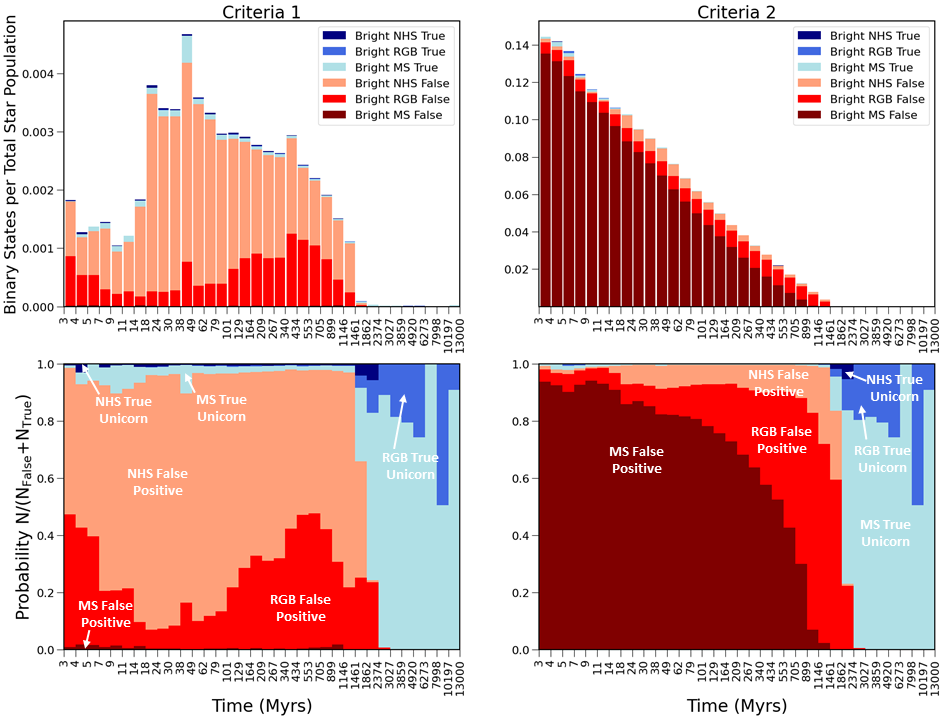}
     \caption{Results of stellar binary evolution modelling under a Hosek IMF, focusing on the probabilities of false positives (red shades) and true Unicorns (blue shades) within the population. \textit{Top left:} shows the rate at which these objects appear at different time intervals as a percentage of the overall population of $5\times10^5$ binaries. This plot utilizes Criteria 1, wherein we require the dark object to be more massive than the bright object, and assumes L$_{\mathrm{Dark}}$<L$_{\mathrm{Bright}}$ by any amount. \textit{Bottom left:} provides the false positive and true Unicorn probabilities within our population of interest under Criteria 1. This gives the likelihood that any given binary with a bright star and a more massive dark object greater than $1.242$~M$_\odot$ forms a true Unicorn. \textit{Top right:} shows the rate at which these objects appear as a percentage of overall population under Criteria 2, which is a looser criteria wherein we do not require the dark object to be more massive than the bright object. \textit{Bottom right:} shows the true Unicorn and false positive rates under Criteria 2. Acronyms refer to:  BH=Black Hole, MS=Main Sequence Star, NHS=Naked Helium Star, NS=Neutron Star, RGB=Red Giant.}
     \label{tab:FigureB1}
 \end{figure*}

Figure \ref{tab:FigureB1} shows the results of our stellar evolution models for the population evolved under a Hosek IMF \citep{Hosek2019}. This gives the likelihood that different objects of interest are either a true Unicorn or false positive at different time intervals. We define true Unicorns as a NS or BH orbited by a MS, RGB, or NHS. False positives are defined as a pair of stars (MS/MS, RGB/MS, RGB/RGB, NHS/MS, or NHS/RGB) which mimic true Unicorns in mass and luminosity as did the binaries observed in \citet{Jayasinghe2021Unicorn} and \citet{Jayasinghe2022Giraffe}.

\bsp	
\label{lastpage}
\end{document}